\documentclass[aps,prc,reprint,showpacs,floatfix,superscriptaddress]{revtex4-1}

\usepackage{amsfonts,amssymb,amsmath}
\usepackage{graphicx}
\usepackage{dcolumn}
\usepackage{bm}
\usepackage[dvips]{color}
\usepackage[colorlinks=true,linkcolor=blue,citecolor=blue,urlcolor=blue]{hyperref}
\hypersetup{
  pdftitle={Nucleon-nucleon potentials in phase-space representation},%
  pdfauthor={Hans Feldmeier <h.feldmeier@gsi.de>}} 

\usepackage[english]{babel}

\begin{document}
\newcolumntype{.}{D{.}{.}{-1}}

\newcommand{\bra}[1]{\big< \,{#1}\, \big| }
\newcommand{\ket}[1]{\big| \,{#1}\, \big> }
\newcommand{\braa}[1]{{}_{a\!}\big< \,{#1}\, \big| }
\newcommand{\keta}[1]{\big| \,{#1}\, \big>{}_{\!a} }
\newcommand{\qstate}[2]{Q^{(#2)};J^\pi M #1}
\newcommand{\Q}[2]{Q^{(#1)};J^\pi #2}
\newcommand{\braket}[2]{\big< \,{#1}\, \big| \,{#2}\, \big> }
\newcommand{\braketa}[2]{\phantom{\big<}_a\!\big< \,{#1}\, \big| \,{#2}\, \big>\!_a }
\newcommand{\expect}[1]{\big< \, {#1} \, \big>}

\newcommand{\matrixe}[3]{\big< \,{#1}\, \big| \,{#2}\, \big| \,{#3}\, \big> }


\newcommand{\op}[1]{\bm{#1}}
\newcommand{\vek}[1]{\!\vec{\,#1}}
\newcommand{\e}{\mathrm{e}}
\renewcommand{\i}{\mathrm{i}}
\newcommand{\ps}[1]{{#1}_{\mathrm{ps}}}
\newcommand{\ndot}{\!\cdot\!}
\newcommand{\se}{\!=\!}
\newcommand{\ClebschGordan}[6]{\left< \begin{array}{cc} #1 & #3 \\ #2 & #4 \end{array} \right|\left. \begin{array}{c} #5 \\ #6 \end{array} \right>}

\title{Nucleon-nucleon potentials in phase-space representation}

\author{H.~\surname{Feldmeier}}
\affiliation{Frankfurt Institute for Advanced Studies, Max-von-Laue-Stra{\ss}e 1, 60438 Frankfurt, Germany}
\affiliation{GSI Helmholtzzentrum f\"ur Schwerionenforschung GmbH, Planckstra{\ss}e 1, 64291
Darmstadt, Germany}

\author{T.~\surname{Neff}}
\affiliation{GSI Helmholtzzentrum f\"ur Schwerionenforschung GmbH, Planckstra{\ss}e 1, 64291
Darmstadt, Germany}
\author{D.~\surname{Weber}}
\affiliation{GSI Helmholtzzentrum f\"ur Schwerionenforschung GmbH, Planckstra{\ss}e 1, 64291
Darmstadt, Germany}
\affiliation{ExtreMe Matter Institute EMMI and Research Division, Planckstra{\ss}e 1, 64291
Darmstadt, Germany} 

\date{\today}
\graphicspath{{img/}}

\newcommand{\noteTN}[1]{{\color{red}[TN: #1]}}

\begin{abstract}
A phase-space representation of nuclear interactions, which depends on the distance $\vek{r}$ and 
relative momentum $\vek{p}$ of the nucleons, is presented. A method is developed that permits to 
extract the interaction $V(\vek{r},\vek{p})$ 
from antisymmetrized matrix elements given in a spherical basis with 
angular momentum quantum numbers, either in momentum or coordinate space representation. 
This representation visualizes in an intuitive way the non-local behavior introduced by
cutoffs in momentum space or renormalization procedures that are used to adapt the interaction to 
low momentum many-body Hilbert spaces, as done in the unitary correlation operator method
or with the similarity renormalization group.  It allows to develop intuition about
the various interactions and illustrates how the softened interactions reduce the short-range repulsion 
in favor of non-locality or momentum dependence while keeping the scattering phase shifts invariant.
It also reveals that these effective interactions can have undesired complicated momentum dependencies at momenta around and above the Fermi momentum.  Properties, similarities and differences of the phase-space representations of the Argonne and the N3LO chiral potential, and their UCOM and SRG derivatives are discussed.
\end{abstract}

\maketitle

\section{Introduction}

Realistic nucleon-nucleon (NN) potentials, such as the Argonne V18 \cite{wiringa95}, 
the Bonn CD potential \cite{machleidt01},
or potentials from chiral perturbation theory 
\cite{entem03,epelbaum02,epelbaum06,gerzelis13,gerzelis14,epelbaum09,lutz00,gasparyan13},
describe experimental two-nucleon data with high accuracy. These NN interactions induce short-range 
and tensor correlations in nuclear many-body states, which are difficult to treat in nuclear many-body
methods.  To derive soft effective interactions for nuclear \textit{ab initio} calculations, 
one often applies transformation techniques, such as the unitary correlation operator method (UCOM)
\cite{ucom98,ucom03,ucom04,ucom05,ucom06,ucom10} 
or the similarity renormalization group (SRG) \cite{Bogner20031,Bogner2003265}
that maintain the phase shifts of the original interaction. 
Consequently, there exists a series of effective realistic interactions, which all succeed in 
describing two-nucleon data with the same precision but differ in their momentum-dependence and off-shell 
behavior. They also lead to different induced three- and many-body interactions.

Many of these effective realistic NN interactions are formulated in a partial wave matrix representation
in momentum space, which is not very suitable for visualizing position space properties of the potential. 
In this article we present a phase-space representation for NN interactions, which depends on relative 
distance and relative momentum. 
This allows to demonstrate  momentum space properties as induced by the
softening of the potential or non-localities originating from having different potentials in 
different partial waves. 
At the same time it also allows to study position space properties like the short-range 
repulsion and its renormalization or 
the effect of the SRG transformation on the long-range part. 
One also sees effects in position space originating from cutoffs in momentum space.
A ``local projection'' of the interaction can be obtained by setting the relative momentum to zero.

This paper extends the work by Wendt, Furnstahl and Ramanan \cite{wendt12}
who investigated local projections of effective nuclear interactions. 
It is recommended to consult this paper for further information as we will not repeat their
findings that also apply to this work.

Another motivation for the phase-space representation is that an intuitive and descriptive picture 
of the interaction is also helpful in creating an operator
based representations for many-body methods that work with wave packets \cite{feldmeier00}
like antisymmetrized molecular dynamics \cite{kanadaenyo95,kanadaenyo01,kanadaenyo12} 
or fermionic molecular dynamics
\cite{fmd90,fmd04,fmd04b,fmd11} because their basis states possess no angular momentum
quantum numbers. 
It also helps to find a meaningful set of operators for representing a given set of
matrix elements \cite{weber14}.

In Sec.~\ref{sec:PhaseSpaceRepresentation} we introduce the phase-space representation proposed by 
Kirkwood \cite{PhysRev.44.31} for nuclear potentials and discuss basic properties of this representation 
for potentials with linear and quadratic momentum and angular momentum dependencies. 
The method to deduce the phase-space representation from partial wave momentum matrix elements 
of the interaction is also presented in this section. 
In Sec.~\ref{sec:NNpotentials} we show the phase-space representation of different effective realistic 
NN interactions and discuss characteristic properties of the considered interactions and how 
UCOM and SRG transformations are reflected in the phase-space representation.

\section{Phase-space Representation \label{sec:PhaseSpaceRepresentation}}

\subsection{Definition \label{sec:Definition}}
We use the phase-space representation introduced by Kirkwood \cite{PhysRev.44.31} to investigate the momentum dependence of NN potentials. The phase-space distribution $\ps{f}(\vek{r},\vek{p})$ for a density operator $\op{\rho}$ and the representation $\ps{O}(\vek{r},\vek{p})$ of an operator $\op{O}$ are defined as
\begin{align}
\ps{f}(\vek{r},\vek{p})&=\bra{\vek{r}}\op{\rho}\ket{\vek{p}}\braket{\vek{p}}{\vek{r}} \label{eqn:phdensity}\\
\ps{O}(\vek{r},\vek{p})&=\left(2\pi\right)^{3}\bra{\vek{r}}\op{O}\ket{\vek{p}}\braket{\vek{p}}
{\vek{r}}\label{eqn:phoperator},
\end{align}
such that the quantum expectation value is given by
\begin{equation}
\expect{\op{O}}=\mathrm{Tr}\left(\op{\rho}\,\op{O}\right)=
\int\mathrm{d}^3r\,\mathrm{d}^3p\,\ps{f}^*(\vek{r},
\vek{p})\ \ps{O}(\vek{r},\vek{p}).
\end{equation}

Note that $\ps{f}(\vek{r},\vek{p})$ and $\ps{O}(\vek{r},\vek{p})$ can be complex
but the expectation value of a hermitean operator is of course real.
The Kirkwood representation is similar to the Wigner representation \cite{Wignerrepresentation}, 
but is easier to evaluate for potentials given in a partial wave decomposition. 

Throughout the paper operators in Hilbert space are denoted by bold characters.

To illustrate the phase-space representation and to train imagination and intuition, 
we first discuss the Kirkwood representation of potentials with simple momentum dependencies.

\subsection{Examples \label{sec:Examples}}
For a local potential $\op{V}^{loc}\!=V(\op{r})$ the phase-space representation is easily calculated by 
applying $\op{V}^{loc}$ to the bra $\bra{\vek{r}}$ in Eq.~(\ref{eqn:phoperator}) as 
\begin{eqnarray}
\ps{V}^{loc}(\vek{r},\vek{p})=\left(2\pi\right)^3V(r)\braket{\vek{r}}{\vek{p}}
\braket{\vek{p}}{\vek{r}}=V(r),\label{eq:loc}
\end{eqnarray}
where we used $\braket{\vek{r}}{\vek{p}}=(2\pi)^{-3/2}{\rm e}^{\i\vek{r}\cdot\vek{p}}$.
Thus, the phase-space representation of a local potential  has the same functional dependence on
the relative distance $\vek{r}$ as the operator and does not depend on the relative momentum $\vek{p}$.

Likewise any operator $\op{V}^{mom}\!=V(\op{p})$ depending only on $\op{p}$ can be applied to the ket 
$\ket{\vek{p}}$ in Eq.~(\ref{eqn:phoperator}) to yield
\begin{eqnarray}
\ps{V}^{mom}(\vek{r},\vek{p})=V(p)\, .
\end{eqnarray}

The phase-space representation of a potential with quadratic momentum dependence like 
$\op{V}^{p2}=\frac{1}{2}\left(\vek{\op{p}}^{2}V(\op{r})+V(\op{r})\,\vek{\op{p}}^{2}\right)$
is given by
\begin{equation}
\ps{V}^{p2}(\vek{r},\vek{p})=V(r)\,p^2-\frac{1}{2}V''(r)-\frac{V'(r)}{r}-\i\frac{V'(r)}{r} \,
\vek{r}\ndot\vek{p}, \label{eq:psp2}
\end{equation}
where $r=|\vek{r}|$ and $p=|\vek{p}|$.
Deriving the phase-space representation of $V(\op{r})\,\vek{\op{p}}^{2}$ by means of 
Eq.~(\ref{eqn:phoperator}) is straightforward and just gives $V(r)\,p^2$. The other term 
$\vek{\op{p}}^{\,2}V(\op{r})$, however, requires to commute $V(\op{r})$ and $\vek{\op{p}}^{\,2}$:
\begin{equation}  \label{eq:psp3}
\vek{\op{p}}^{2}V(\op{r})=V(\op{r})\,\vek{\op{p}}^{2}-V''(\op{r})-2\frac{V'(\op{r})}{\op{r}}
   -2\frac{V'(\op{r})}{\op{r}}\ \i\,\vek{\op{r}}\ndot\vek{\op{p}} \, .
\end{equation}
 
This creates the terms with derivatives of $V(r)$ in Eq.~(\ref{eq:psp2}) and the phase-space 
representation contains, besides the expected quadratic momentum-dependence, additional local 
contributions and a term with the scalar product $\i\,\vek{r}\ndot\vek{p}$. As $\vek{p}$ is odd 
under a time reversal operation the appearance of $\i=\sqrt{-1}$ is expected because $\op{V}^{p2}$ 
is even under time reversal.

For interactions containing the relative angular momentum $\vek{\op{L}}=\vek{\op{r}}\times\vek{\op{p}}$
in the coordinate space part, like the spin-orbit force, the phase-space representation of 
$\op{V}^{L}=V(\op{r})\,\vek{\op{L}}$ is given by
\begin{eqnarray}
\ps{V}^{L}(\vek{r},\vek{p})=V(r)\ \vek{r}\times\vek{p}\ . \label{eq:psL}
\end{eqnarray}

For potentials with a quadratic angular momentum dependence, 
$\op{V}^{L2}=V(\op{r})\,\vek{\op{L}}\phantom{}^{2}$, 
one finds the phase-space representation
\begin{eqnarray}
\ps{V}^{L2}(\vek{r},\vek{p})&=&V(r)\left(\vek{r}\times\vek{p}\right)^2+2\i V(r)\,\vek{r}\ndot\vek{p}
\nonumber \\
&=&V(r)r^2p^2+2\i V(r) \,\vek{r}\ndot\vek{p}-V(r)(\vek{r}\ndot\vek{p})^2. \label{eq:psL2}
\end{eqnarray}
In this paper we discuss  only the $S=0$ channel. The treatment of spin-orbit or tensor interactions 
will be subject of further studies. 

For $S=0$ the coordinate part of the interaction $\op{V}=V(\op{\vek{r}},\op{\vek{p}}\,)$ has to be 
invariant under rotations and thus it is, like its phase-space representation, a function of the 
absolute values $r$ and $p$, and the scalar product $\vek{r}\cdot\vek{p}$. 
For illustration purposes it is convenient to expand the phase-space 
representation in powers of $\hat{r}\cdot\hat{p}$ or in terms of Legendre polynomials 
$P_\Lambda$ as
\begin{equation}
 \ps{V}(\vek{r},\vek{p})=\ps{V}(r,p,\vek{r}\ndot\vek{p})=
\sum_{\Lambda=0}^\infty  \ps{V}^\Lambda(r,p)\,\i^\Lambda P_\Lambda(\hat{r}\ndot\hat{p}).\label{eq:expansion}
\end{equation}
and show two-dimensional plots for $\ps{V}^\Lambda(r,p)$ for different $\Lambda$.
$\hat{x}$ denotes the unit vector in the direction of $\vek{x}$.

The operator relation
\begin{equation}  \label{eq:psp4}
\vek{\op{p}}^{2}\ \i\,\vek{\op{r}}\ndot\vek{\op{p}}=
\i\,\vek{\op{r}}\ndot\vek{\op{p}}\ \vek{\op{p}}^{2}+2\,\vek{\op{p}}^{2}
\end{equation}
together with Eq.~\eqref{eq:psp3} implies that any power of $\vek{\op{p}}^{2}$ combined with any
radial dependence $V(\op{r})$ leads in the Kirkwood phase-space representation to
terms proportional to $({\hat{r}\ndot\hat{p}})^0$ and $(\hat{r}\ndot\hat{p})$ only, which means
that only $\Lambda=0$ and $\Lambda=1$ contribute.

This is different for powers of $\vek{\op{L}}^{2}$, where for $\big(\vek{\op{L}}^{\,2}\big)^n$
terms up to $\Lambda=2n$ contribute.

The relation between the Legendre polynomials and the directions of $\vek{r}$ and $\vek{p}$ expressed
in terms of spherical harmonics $Y^\Lambda(\hat{r})$ and $Y^\Lambda(\hat{p})$ given by the addition
theorem
\begin{align}
P_\Lambda(\hat{r}\ndot\hat{p})=&\frac{4\pi}{2\Lambda+1} \label{eq:PlY}
            \sum_{\mu=-\Lambda}^{\Lambda} {Y^\Lambda_\mu}^*\!(\hat{r}){Y^\Lambda_\mu}(\hat{p}) \\
        =&\frac{4\pi(-1)^\Lambda}{\sqrt{2\Lambda+1}}
             \left[Y^\Lambda(\hat{r}){Y^\Lambda}(\hat{p})\right]^0
\end{align}
reflects the fact that the $S=0$ potentials are of tensor rank zero in coordinate space.

The coordinate space part $V(\op{r})\,\vek{\op{L}}$ of the spin-orbit interaction is 
a vector or a tensor of rank 1 in
coordinate space and can thus not be written in terms of $P_\Lambda$ but in terms of 
$\left[Y^\Lambda(\hat{r}){Y^\Lambda}(\hat{p})\right]^1$.

Rewriting Eqs.~(\ref{eq:loc})-(\ref{eq:psp2}) and (\ref{eq:psL2}) in terms of the expansion 
in Legendre polynomials leads to
\begin{eqnarray}
\ps{V}^{loc}(\vek{r},\vek{p})&=&V(r)\,P_0(\hat{r}\ndot\hat{p})\label{eq:partloc}\\
\ps{V}^{mom}(\vek{r},\vek{p})&=&V(p)\,P_0(\hat{r}\ndot\hat{p})\label{eq:partmom}\\
\ps{V}^{p2}(\vek{r},\vek{p})&=&\left[V(r)\,p^2-\frac{1}{2}V''(r)-
\frac{V'(r)}{r}\right]P_0(\hat{r}\ndot\hat{p})\nonumber\\
&&-\frac{V'(r)}{r}rp\,\i\,P_1(\hat{r}\ndot\hat{p})\label{eq:partp2}\\
\ps{V}^{L2}(\vek{r},\vek{p})&=&
\frac{2}{3}V(r)\left(rp\right)^2\,P_0(\hat{r}\ndot\hat{p})+
                                       2V(r)\,rp\,\i\,P_1(\hat{r}\ndot\hat{p})\nonumber \\
&&+\frac{2}{3}V(r)\left(rp\right)^2\,\i^2\,P_2(\hat{r}\ndot\hat{p})\ .\label{eq:partL2}
\end{eqnarray}

\begin{figure*}[ht!]
\centering
\includegraphics[width=0.8\textwidth]{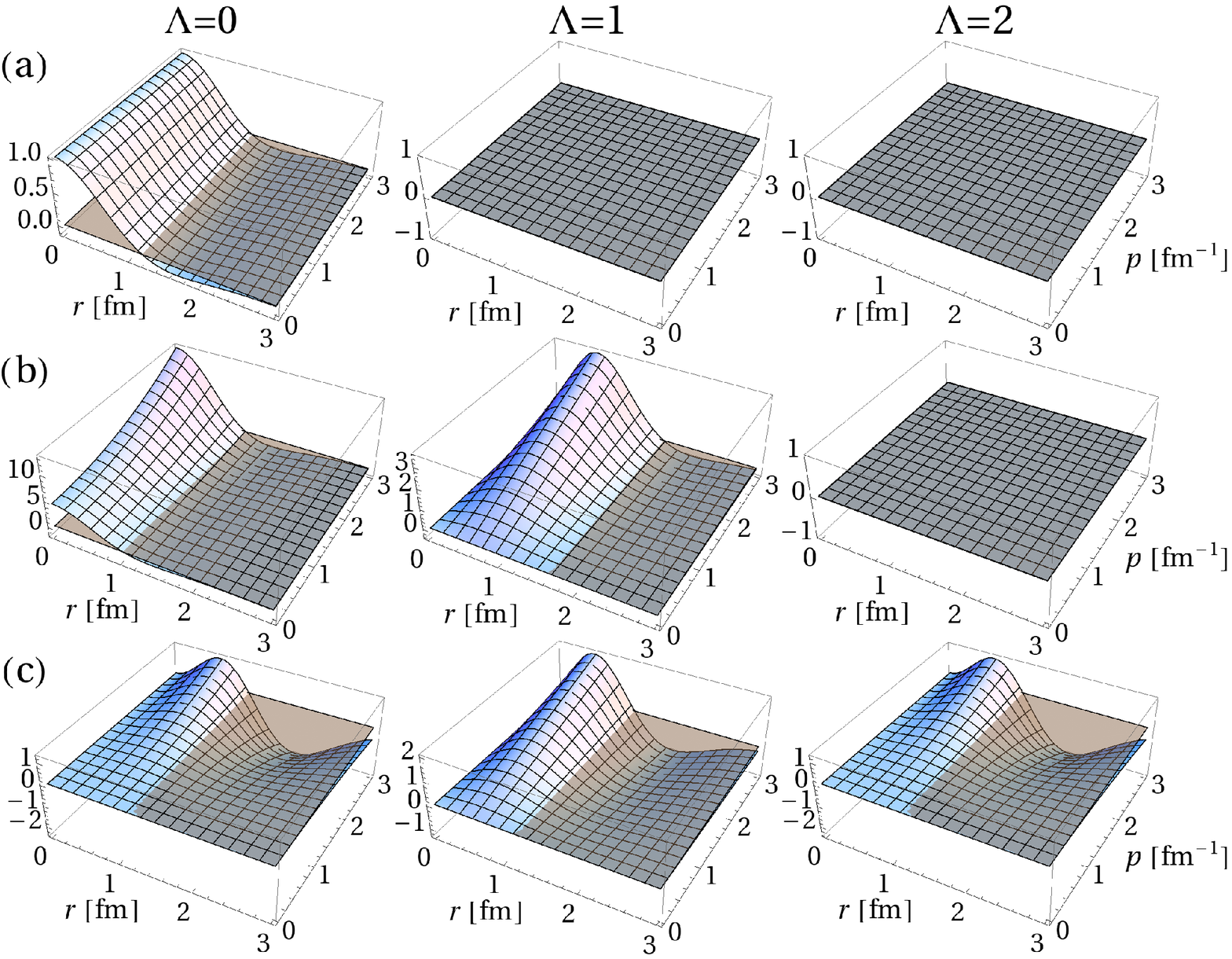}
\caption{\label{fig:vph}(Color online) 
Phase-space representation $V_{ps}^{\Lambda}(r,p)$ in arbitrary 
units for (a)  $\op{V}^{loc}\!=\!V(\op{r})$, (b) $\op{V}^{p2}\!=\!\frac{1}{2}
\left(\vek{\op{p}}^{\,2}V(\op{r})+V(\op{r})\vek{\op{p}}^{\,2}\right)$, 
(c) $\op{V}^{L2}\!=\!V(\op{r})\,\vek{\op{L}}^2$. The radial function is taken to be a sum of gaussians:  $V(r)\!=\!2\e^{-\frac{r^2}{1\,\mathrm{fm}^2}}-\e^{-\frac{r^2}{2\,\mathrm{fm}^2}}$.}
\end{figure*}

To illustrate the phase-space representation of potentials with short-range repulsion and 
long-range attraction we show in Fig.~\ref{fig:vph} the components $\ps{V}^\Lambda(r,p)$ of a local, 
a quadratic momentum dependent, and a quadratic angular momentum dependent
potential for $\Lambda=0,\,1,\,2$. The phase-space representation of the local potential is independent 
of $p$ and has no angular dependence, so that only $\Lambda=0$ contributes. 
The quadratic momentum potential shows a $p^2$-dependence for $\Lambda=0$, and a $\Lambda=1$ component, 
which is proportional to $p$ and originates from the  scalar product $\vek{r}\cdot\vek{p}$ in
Eq.~(\ref{eq:psp2}) reflecting the fact that $\vek{\op{r}}$ and $\vek{\op{p}}$ do not commute. 

The given examples contain terms up to $\Lambda=2$. All higher $\Lambda$-contributions vanish. 
Potentials with higher powers in $\vek{\op{L}}\phantom{}^{2}$, like those defined in each $L$-channel
independently or a SRG transformation (shown later), 
create contributions also for higher $\Lambda$ or higher powers in $\hat{r}\ndot\hat{p}$.
Thus, the terms $\ps{V}^\Lambda(r,p)$ of the expansion Eq.~(\ref{eq:expansion}) visualize in an 
intuitive way the radial and momentum dependence of a nuclear interactions.

The local projection of a non-local potential as defined in Eq.~(3) of Ref.~\cite{wendt12} 
is simply $\ps{V}(\vek{r},\vek{p}\!=\!0)$. This follows immediately from Eq.~\eqref{eqn:phoperator}:
\begin{align}\label{eq:localproj}
\ps{V}(\vek{r},\vek{p}\!=\!0) &= \ps{V}^{\Lambda=0}(\vek{r},\vek{p}\!=\!0)\nonumber \\
&= (2\pi)^3\!\!\int\!\! d^3 r'   \bra{\vek{r}}\op{V}\ket{\vek{r}'}
                          \braket{\vek{r}'}{\vek{p}\!=\!0}\braket{\vek{p}\!=\!0}{\vek{r}}\nonumber \\
   &= \int d^3 r' \, \bra{\vek{r}}\op{V}\ket{\vek{r}'}  \, .
\end{align}

For a potential given in operator representation its phase-space representation 
$\ps{V}(\vek{r},\vek{p})$ can be calculated by means of 
Eq.~(\ref{eqn:phoperator}). The UCOM transformed Argonne potential, for example, contains  for $S=0$  a 
central, a quadratic momentum dependent, and a quadratic angular momentum part, so that its phase-space 
representation can be calculated directly from Eqs.~(\ref{eq:partloc}) and 
(\ref{eq:partp2})-(\ref{eq:partL2}). However, many NN potentials are only available in matrix element 
representation. In this case, we have to rewrite Eq.~(\ref{eqn:phoperator}) in terms of the available 
interaction matrix elements. A complication arises due to the fact that the matrix elements are 
calculated with antisymmetrized two-body states, which will be addressed in the next section. 

\subsection{Phase-space representation for (anti-)symmetric coordinate space} 
\label{sec:antisymmPSR}
In the nuclear case the particle pair has the intrinsic degrees of freedom spin $S=0,1$ and
isospin $T=0,1$. As the two-particle states have to be antisymmetric under particle exchange
$(\,(\vek{r}_1-\vek{r}_2)\!=\!\vek{r}\longrightarrow(\vek{r}_2-\vek{r}_1)\!=\!-\vek{r}\,)$
the coordinate space part of the two-body state can be either symmetric or antisymmetric. 
Therefore the Kirkwood representation \eqref{eqn:phoperator} has to be adapted. For that let us
introduce the basis states
\begin{align}
\ket{\vek{r};S,M_S;T,M_T} \\
\ket{\vek{p};S,M_S;T,M_T}
\end{align}
with
\begin{align}
\braket{\vek{r}}{\vek{p};S,M_S;T,M_T}=& \nonumber \\
   (2\pi)&^{-3/2}\ {\e}^{\i\vek{p}\cdot\vek{r}}\ \ket{S,M_S;T,M_T}\, 
\end{align}
and the corresponding antisymmetrized states by
\begin{widetext}
\begin{align}
\keta{\vek{r};S,M_S;T,M_T}=& 
  \frac{1}{\sqrt{2}}\big( \ket{\vek{r};S,M_S;T,M_T}-(-1)^{T+S}  \ket{-\vek{r};S,M_S;T,M_T} \big) \\
\keta{\vek{p};S,M_S;T,M_T}=& 
   \frac{1}{\sqrt{2}}\left( \ket{\vek{p};S,M_S;T,M_T}-(-1)^{T+S}  \ket{-\vek{p};S,M_S;T,M_T}\right)
\end{align}
with
\begin{align}
\braket{\vek{r}}{\vek{p};S,M_S;T,M_T}_{\!a}= {(2\pi)^{-3/2}} \frac{1}{\sqrt{2}}
    \left( {\e}^{\i\vek{p}\cdot\vek{r}}-(-1)^{T+S} {\e}^{-\i\vek{p}\cdot\vek{r}}\right)
    \ket{S,M_S;T,M_T} \, .
\end{align}
One sees that for $S+T$ being an even number the wave function is proportional to 
$\sin(\vek{p}\cdot\vek{r})$, which means being antisymmetric under particle exchange (negative parity), 
and for odd $S+T$ it is proportional to $\cos(\vek{p}\ndot\vek{r})$, which means being symmetric 
(positive parity).

The matrix element of an interaction $\op{V}$ calculated with antisymmetric basis states 
\begin{align}
\braa{\vek{r};S,T}\op{V}\keta{\vek{p};S,T}=& 
 \bra{\vek{r};S,T}\op{V}\ket{\vek{p};S,T}
                  -(-1)^{T+S} \bra{\vek{r};S,T}\op{V}\ket{-\vek{p};S,T} 
\end{align}
consists of a direct and an exchange term. Please note that we omitted the $M$ quantum numbers because
the nuclear interaction $\op{V}$ is invariant under rotations and diagonal in $S$ and $T$.

In analogy to the Kirkwood representation in Eq.~\eqref{eqn:phoperator} we define a spin-isospin-dependent 
phase-space representation by
\begin{align}
\ps{\mathcal{V}}^{S,T}(\vek{r},\vek{p})&=(2\pi)^3\braa{\vek{r};S,T}\op{V}\keta{\vek{p},S,T}
\braket{\vek{p}}{\vek{r}}   \label{eq:ps-antisymm}\\ \nonumber
&= (2\pi)^3\Big(\bra{\vek{r};S,T}\op{V}\ket{\vek{p};S,T}
                  -(-1)^{T+S} \bra{\vek{r};S,T}\op{V}\ket{-\vek{p};S,T}\Big)\braket{\vek{p}}{\vek{r}}\\
&=\ps{V}^{S,T}(\vek{r},\vek{p})-(-1)^{T+S} {\e}^{-2\i\vek{r}\cdot\vek{p}}
                             \ \ps{V}^{S,T}(\vek{r},-\vek{p}) \, .
\label{eqn:rpST}
\end{align}
\end{widetext}
Our intuition is trained for $\ps{V}^{S,T}(\vek{r},\vek{p})$
and not for $\ps{\mathcal{V}}^{S,T}(\vek{r},\vek{p})$. Therefore, we have to develop a method to
extract the direct term $\ps{V}^{S,T}(\vek{r},\vek{p})$ from $\ps{\mathcal{V}}^{S,T}(\vek{r},\vek{p})$.

Before proceeding we would like to introduce some simplifications.
As we will consider in this paper only the case $S=0$, from now on we will use a shortened notation 
omitting the quantum number $S$ and implying that the total spin equals the orbital angular momentum, $J=L$.

Contrary to the situation without spin and isospin as discussed in section~\ref{sec:Examples}
the phase-space representation of a purely local and isospin independent potential $V(\op{r})$ already features a momentum dependence and is different for $T=0$ and $T=1$ due to the exchange term. Using the
definition given in Eq.~\eqref{eq:ps-antisymm} one gets for this case 
\begin{eqnarray}
\ps{\mathcal{V}}^T(\vek{r},\vek{p})=
\ps{V}(\vek{r})-(-1)^T\,{\e}^{-2\i\vek{r}\cdot\vek{p}}\ \ps{V}(\vek{r})\, .\label{eq:matrixeq1}
\end{eqnarray}
Thus, for a local potential the direct term $\ps{V}(\vek{r})$ does not depend on momentum but the
exchange part $(-1)^T\,{\e}^{-2\i\vek{r}\cdot\vek{p}}\ps{V}(\vek{r})$ is momentum and 
isospin dependent.

In the general case where the operator depends on isospin, distance and relative momentum the
phase-space representation takes the form Eq.~\eqref{eqn:rpST}
\begin{eqnarray}
\ps{\mathcal{V}}^T(\vek{r},\vek{p})=\ps{V}^T(\vek{r},\vek{p})
                  -(-1)^T \,{\e}^{-2\i\vek{r}\cdot\vek{p}}\,\ps{V}^T(\vek{r},-\vek{p})\, .
\end{eqnarray}
As in Eq.~\eqref{eq:expansion} this is expanded in terms of Legendre polynomials
$P_{\Lambda'}(\hat{r}\ndot\hat{p})$
\begin{eqnarray}
 \ps{\mathcal{V}}(\vek{r},\vek{p})=
\sum_{\Lambda'=0}^\infty  \ps{\mathcal{V}}^{T,\Lambda'}(r,p)\ \i^{\Lambda'} 
                                      P_{\Lambda'}(\hat{r}\ndot\hat{p})\, ,
\label{eq:expansion-antisym}
\end{eqnarray}
where one obtains for each $\Lambda'$
\begin{align}
\ps{\mathcal{V}}^{T,\Lambda'}(r,p)=& \nonumber \\
           \ps{V}^{T,\Lambda'}(r,p)&-(-1)^T\sum_{\Lambda=0}^\infty  \ps{V}^{T,\Lambda}(r,p)\
    C_{\Lambda \Lambda'}(r,p)\, ,
\label{eq:matrixeq2}
\end{align}
with the coefficients
\begin{align}
C_{\Lambda \Lambda'}(r,p)& = \nonumber \\
\sum_{L} (-\i)&^{L+\Lambda'+\Lambda}\, j_L(2rp)\,(2L+1) 
\ClebschGordan{L}{0}{\Lambda}{0}{\Lambda'}{0}^2 
\end{align}
The summation over $L$ is limited to the range $|\Lambda-\Lambda'|\le L \le \Lambda+\Lambda'$
and the Clebsch-Gordan coefficients vanish for $L+\Lambda'+\Lambda$ being odd.
The coefficients originate from the partial wave decomposition of the plane wave 
$\e^{-2\i\vek{r}\cdot\vek{p}}$
in Eq.~(\ref{eq:matrixeq1}). The detailed derivation is given in the appendix.

With help of these coefficients, which depend on $r$ and $p$, we define the matrix
\begin{align}
A^T_{\Lambda\Lambda'}(r,p)=\delta_{\Lambda\Lambda'}-(-1)^T\,C_{\Lambda\Lambda'}(r,p) 
\label{eq:matrixA}
\end{align}
and write Eq.~\eqref{eq:matrixeq2} as
\begin{align}
\ps{\mathcal{V}}^{T,\Lambda'}\!(r,p)= \sum_{\Lambda=0}^\infty 
           \ps{V}^{T,\Lambda}(r,p)\ A^T_{\Lambda\Lambda'}(r,p)\, .
\label{eq:matrixeq3}
\end{align}
Under assumptions discussed in the following section the inversion of $A^T_{\Lambda\Lambda'}(r,p)$ is 
possible and the phase-space representation $\ps{V}^{T,\Lambda}(r,p)$ of the direct 
term can be extracted from the representation $\ps{\mathcal{V}}^{T,\Lambda'}\!(r,p)$
obtained in the antisymmetrized case.\\

%
\subsection{Partial wave matrix elements} \label{sec:partialwave}
In this work we also want to visualize potentials that are given only in form of their momentum space 
partial wave matrix elements $_a\bra{k(LS)JM;TM_T}\op{V}\ket{k'(L'S)JM;TM_T}_a$, 
with the relative momentum quantum numbers $k,k'$, 
relative angular momenta $L,L'$, spin $S$, isospin $T,M_T$, and total angular momentum quantum numbers $J$ 
and $M$. Since $\op{V}$ is invariant under rotation, the matrix elements do not depend on $M$  
and the index will be omitted. Isospin symmetry is also assumed. 
As we consider here only $S=0$ we abbreviate the notation to 
$\braa{k;L,T}\op{V}\keta{k';L,T}$, where we have already used that for $S=0$ the interaction $\op{V}$ 
is diagonal in $L$. The overlap of the partial wave states with position and momentum eigenstates are
\begin{widetext} 
\begin{align}
\braket{\vek{r}}{k;LM,T}\!_a 
=&\left\{ 
\begin{array}{cl} 
\sqrt{\frac{2}{\pi}}\,\i^L j_L(kr) Y^L_M(\hat{r}) \ket{S\!=\!0,T} & \mathrm{for}\ T+L\ \mathrm{odd}\\ 
      0 &  \mathrm{for}\ T+L\ \mathrm{even}
\end{array} \right.   \label{eq:expansion-rk}\\
\braket{\vek{p}}{k;LM,T}\!_a
=& \left\{   
\begin{array}{cl} 
\frac{\ \,\delta(p-k)}{p^2}\, Y^L_M(\hat{p})\, \ket{S\!=\!0,T} \hspace*{2em} & \mathrm{for}\ T+L\ 
                                                                                         \mathrm{odd}\\
      0 &  \mathrm{for}\ T+L\ \mathrm{even} \ ,
\end{array} \right. \label{eq:expansion-pk}
\end{align}
where $j_L$ is a spherical Bessel function and $Y^L_M$ a spherical harmonic. By inserting
\begin{eqnarray}
\op{V}_{S=0,T}=\sum_{L,M}\int_0^\infty\!\!\!\mathrm{d}k\,k^2\int_0^\infty\!\!\!\mathrm{d}k'\,k'^2\ 
\keta{k;LM,T}\braa{k;L,T}\op{V}\keta{k';L,T}\,\braa{k';LM,T}
\end{eqnarray}
in Eq.~(\ref{eq:ps-antisymm}) and using Eqs.~\eqref{eq:expansion-rk} and \eqref{eq:expansion-pk} 
one gets with the antisymmetrized matrix elements in the mixed $r,p$ representation, defined as
\begin{eqnarray} \label{eq:rp-antisym}
\braa{r;L,T}\op{V}\keta{p;L,T}=\int_0^\infty\!\!\!\mathrm{d}k\,k^2\ 
     j_{L}(kr)\,\braa{k;L,T}\op{V}\keta{p;L,T} \, ,
\end{eqnarray}
the antisymmetrized phase-space representation
\begin{eqnarray}
\ps{\mathcal{V}}^{T}(\vek{r},\vek{p})&=&4\pi \e^{-\i\vek{r}\cdot\vek{p}}
\sum_{L} \braa{r;L,T}\op{V}\keta{p;L,T}\, \sum_M \i^L\,Y^{L}_{M}(\hat{r})\,{Y^{L}_{M}}^*\!(\hat{p}) \, .
\label{eq:pw1}
\end{eqnarray}
Using the formulas~\eqref{eq:YYnachY*} of the appendix to combine the spherical harmonics of the expansion 
of $\e^{-\i\vek{r}\cdot\vek{p}}$ with those appearing in Eq.~\eqref{eq:pw1} one obtains the expansion into 
Legendre polynomials
\begin{equation}
\ps{\mathcal{V}}^{T}(\vek{r},\vek{p})=
\sum_{\Lambda'} \ps{\mathcal{V}}^{T,\Lambda'}\!(r,p)\
\i^{\Lambda'} P_{\Lambda'}(\hat{r}\ndot\hat{p}) \, ,
\end{equation}
where the $\Lambda'$ component of the antisymmetrized phase-space representation, 
$\ps{\mathcal{V}}^{T,\Lambda'}\!(r,p)$, is given in terms of antisymmetrized $r,p$ matrix elements 
(Eq.~\eqref{eq:rp-antisym}) as
\begin{eqnarray} \label{eq:VLambda-rp-antisym}
\ps{\mathcal{V}}^{T,\Lambda'}\!(r,p)&=&
\sum_{L,L'} {\i^{L-L'-\Lambda'}} (2L+1)(2L'+1)
\ClebschGordan{L}{0}{L'}{0}{\Lambda'}{0}^2\,j_{L'}(rp)\
\braa{r;L,T}\op{V}\keta{p;L,T} \, .
\end{eqnarray}
\end{widetext}
For a given $\Lambda'$ the summation over $L$ and $L'$ is limited by the triangular condition
$|L-L'|\le\Lambda'\le L+L'$ of the Clebsch Gordan coefficient. 
The $r,p$ matrix elements vanish for large $L$ due to the finite range of the interaction $\op{V}$,
typically $L_{max}=8$.

The antisymmetric phase-space representation Eq.~\eqref{eq:VLambda-rp-antisym} constitutes the 
left hand side of Eq.~\eqref{eq:matrixeq3}. 
We solve for $\ps{{V}}^{T,\Lambda}(r,p)$ by means of the pseudo inverse $\left(A^T\right)^{+}\!(r,p)$
\begin{eqnarray}
\ps{V}^{T,\Lambda}(r,p)=\sum_{\Lambda'=0}^{\Lambda_{max}}
\left(A^T\right)^{+}_{\Lambda \Lambda'}\!(r,p) \ \ps{\mathcal{V}}^{T,\Lambda'}\!(r,p)
\label{eq:pseudoinv}
\end{eqnarray}
to obtain the least-square solution for $\ps{V}^{T,\Lambda}(\vek{r},\vek{p})$
with a minimal deviation  
\begin{equation} \label{eq:D}
D(r,p)=\!\!
\sum_{\Lambda'=0}^{\Lambda_{max}}  \! \left(
\sum_{\Lambda=0}^{\Lambda_V} \ps{V}^{T,\Lambda}(r,p) A^{T}_{\Lambda\Lambda'}(r,p) -
\ps{\mathcal{V}}^{T,\Lambda'}\!(r,p)  \right)^{\!2}
\end{equation}
for each $(r,p)$ pair.

As mentioned before, there is no information for direct matrix elements
$\bra{k;L,T}\op{V}\ket{p;L,T}$ with $L+T$ being an even number,
because in the angular momentum basis the exchange matrix element differs only
by a factor $(-1)^{L+T}$ from the direct one and thus the antisymmetrized 
$\braa{k;L,T}\op{V}\keta{p;L,T}$ vanishes for $L+T$ even.
In other words, antisymmetrization, which is  a projection, throws away information 
contained in $\ps{V}^{T}(\vek{r},\vek{p})$. Therefore the mapping Eq.~\eqref{eq:matrixeq3}
from $\ps{V}^{T}(\vek{r},\vek{p})$ to $\ps{\mathcal V}^{T}(\vek{r},\vek{p})$ is in general not 
reversible.

We circumvent this problem by assuming that the matrix elements
$\braa{k;L,T}\op{V}\keta{p;L,T}$ for different $L$ are not independent arbitrary sets
of numbers but originate from operators $\op{V}$ that have limited powers in
$\op{\vek{L}}\phantom{}^2$. This implies that the expansion in terms
of Legendre polynomials is limited to a certain $\Lambda_{V}$ determined by the
properties of $\op{V}$. It also means that there are correlations among the matrix elements for different $L$.

We saw in Sec.~\ref{sec:Examples} that the phase-space representation 
$\ps{V}^T(\vek{r},\vek{p})$ of a local potential contains only the $\ps{V}^{T,\Lambda=0}(r,p)$ 
contribution. Therefore, if the potential $\op{V}$ would be local, the sum over
$\Lambda$ in Eq.~\eqref{eq:matrixeq3} would have only one term ($\Lambda_{V}=0$) from which all
$\ps{\mathcal V}^{T,\Lambda'}\!(r,p)$ are determined. 
Likewise a potential that is proportional to $\op{\vek{L}}\phantom{}^2$
would need $\Lambda=0,1,2$ or $\Lambda_{V}=2$ to determine the antisymmetrized 
$\ps{\mathcal V}^{T,\Lambda'}\!(r,p)$ for all $\Lambda'$.

Therefore, we apply the following strategy: We choose a cutoff $\Lambda_{max}$ and 
calculate for each $(r,p)$ the pseudo inverse $\left(A^T\right)^{+}_{\Lambda' \Lambda}(r,p)$
for $\Lambda'$ up to $\Lambda_{max}$, and $\Lambda$ up to $\Lambda_V$. We start with
$\Lambda_V=0$ and then increase $\Lambda_V$ successively by one, until the solution for 
$\ps{V}^{T}(\vek{r},\vek{p})$ obtained from Eq.~\eqref{eq:pseudoinv} succeeds to describe 
$\ps{\mathcal{V}}^T(\vek{r},\vek{p})$, in the sense that the residual deviation $D(r,p)$
defined in Eq.~\eqref{eq:D} is small enough.

\begin{figure}[ht!]
\flushleft (a)\\
\ \ \ \includegraphics[width=0.43\textwidth]{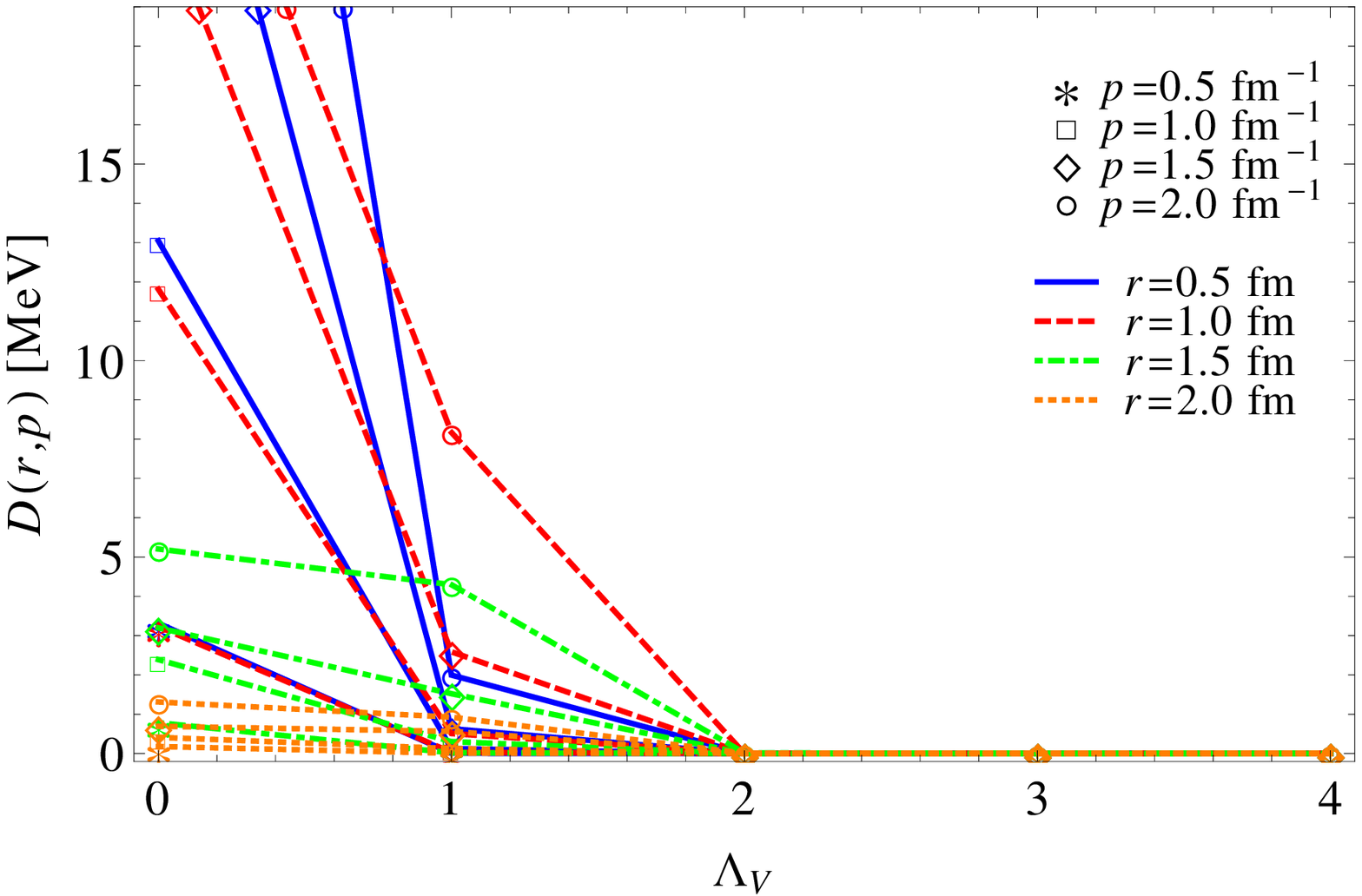}\\
\flushleft (b)\\
\ \ \ \includegraphics[width=0.43\textwidth]{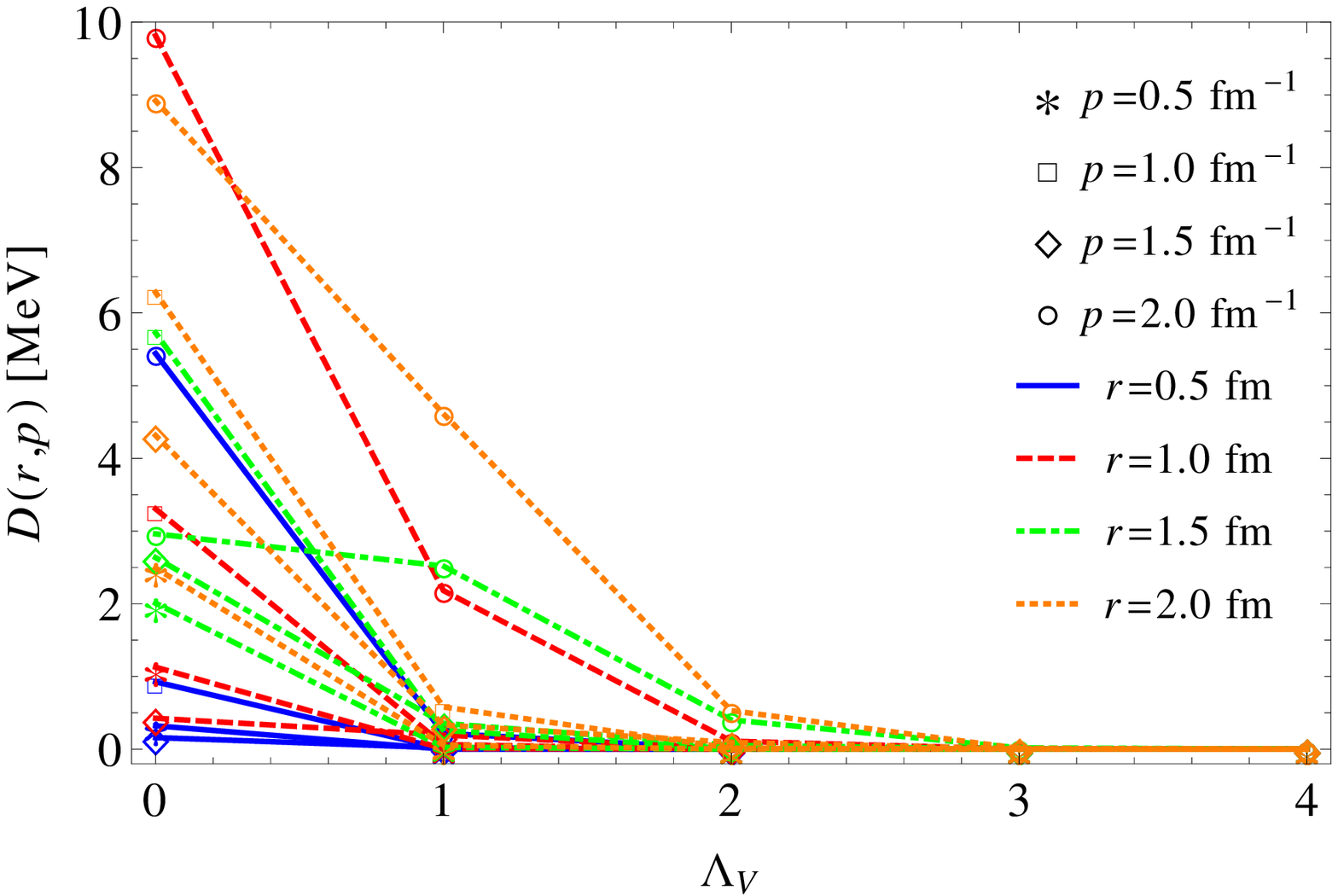}
\caption{\label{fig:res}(Color online) Deviations $D(r,p)$ in the phase-space representation
of the Argonne potential (a) and its SRG ($\alpha=0.2\mathrm{\ fm}^4$) transformation 
(b) for various $r$ and $p$ as a function of $\Lambda_V$.}
\end{figure}

For a local potential this can be fulfilled already for $\Lambda_{V}=0$. 
For the Argonne potential, which has for $S=0$ the form 
$\op{V}^{\mathrm{Arg}}_{S=0,T}=V_{0,T}^C(\op{r})+V^{L2}_{0,T}(\op{r})\,\vec{\op{L}}^{\,2}$, at most 
$\Lambda_V=2$ should be needed (see Eq.~\eqref{eq:partL2}). 
This procedure is illustrated in Fig.~\ref{fig:res}(a), where the deviations $D(r,p)$ for various $r$ and $p$
are plotted as a function of $\Lambda_V$.
It is obvious that no values for $\Lambda_V > 2$ are required as the deviations stay zero and
the phase-space representation does not change anymore. All possible information is already
extracted from the antisymmetric matrix elements. After the SRG transformation, to be discussed
later, one has to sum up to $\Lambda_V=3$ to represent the SRG transformed potential with sufficient 
accuracy. The additional contributions from $\Lambda=4$ are very small even for large $r$ and $p$.
For the SRG transformed N3LO potential the $\Lambda=3$ contributions are larger but again $\Lambda=4$
does not contribute any more. Therefore for all phase-space
representations shown in the following section $\Lambda_V=3$ is used.

%
\section{Phase-space Representation for NN potentials \label{sec:NNpotentials}}
The method discussed in Sec.~\ref{sec:PhaseSpaceRepresentation} is applied to calculate the phase-space 
representations of various effective realistic NN potentials. 
First we display and discuss the $r,p$ matrix elements for each channel $L$ and then combine them,
using the strategy discussed in Sec.~\ref{sec:partialwave}, to generate the
phase-space representation of the interactions for $\Lambda$ up to 3. We investigate
the Argonne and the N3LO chiral potential and effective interactions obtained by UCOM and SRG transformations.

The unitary correlation operator method (UCOM) was introduced in 1998 \cite{ucom98} as an alternative to the then popular G-Matrix approach. Here a unitary
transformation $\op{C}$ is used to imprint explicitly the short range correlations into 
low momentum many-body states $\ket{\Phi}$ 
\begin{equation}
\ket{\widehat{\Phi}}=\op{C}\,\ket{\Phi} \, .
\end{equation}
The unitary transformation is independent of the many-body system, unlike the Pauli projection on 
unoccupied states, or in a Jastrow ansatz, where the many-body states have to be normalized. 
In UCOM the correlations are induced by a similarity transformation that moves 
a pair of particles when they are in the area of the repulsive core from their short distance
$r$ to a larger $R_+(r)$ out of the repulsive region. In this way the wave function 
is depleted at short distances, for details see Ref.~\cite{ucom98}. A many-body matrix element 
of the Hamiltonian calculated
with appropriate correlated states $\ket{\widehat{\Phi}}$ is formally the same as a matrix 
element of an effective Hamiltonian calculated with the corresponding low momentum states:
\begin{align}\label{eq:UCOM}
\matrixe{\widehat{\Phi}}{\op{H}}{\widehat{\Phi}'}& =
    \matrixe{\Phi}{\op{C}^\dagger\op{H}\op{C}}{\Phi'}
    = \matrixe{\Phi}{\op{H}_\mathit{eff}}{\Phi'}\nonumber\\
    & =\matrixe{\Phi}{\op{T}+\op{V}_{U\!C\!O\!M}}{\Phi'} + \cdots \ .
\end{align}
The correlation operator $\op{C}$ is a many-body operator and thus $\op{H}_\mathit{eff}$
contains induced many-body interactions. The UCOM interaction is defined as the two-body part of the transformed Hamiltonian.
The dots in Eq.~\eqref{eq:UCOM} indicate the existence of induced and genuine many-body 
matrix elements, which are absent in the two-body space considered in this paper.
As the shift is only performed at short distances the phase shifts are invariant under this transformation.

\begin{figure*}[t!]
\centering
\begin{minipage}[c]{0.9\textwidth}
\includegraphics[width=0.9\textwidth]{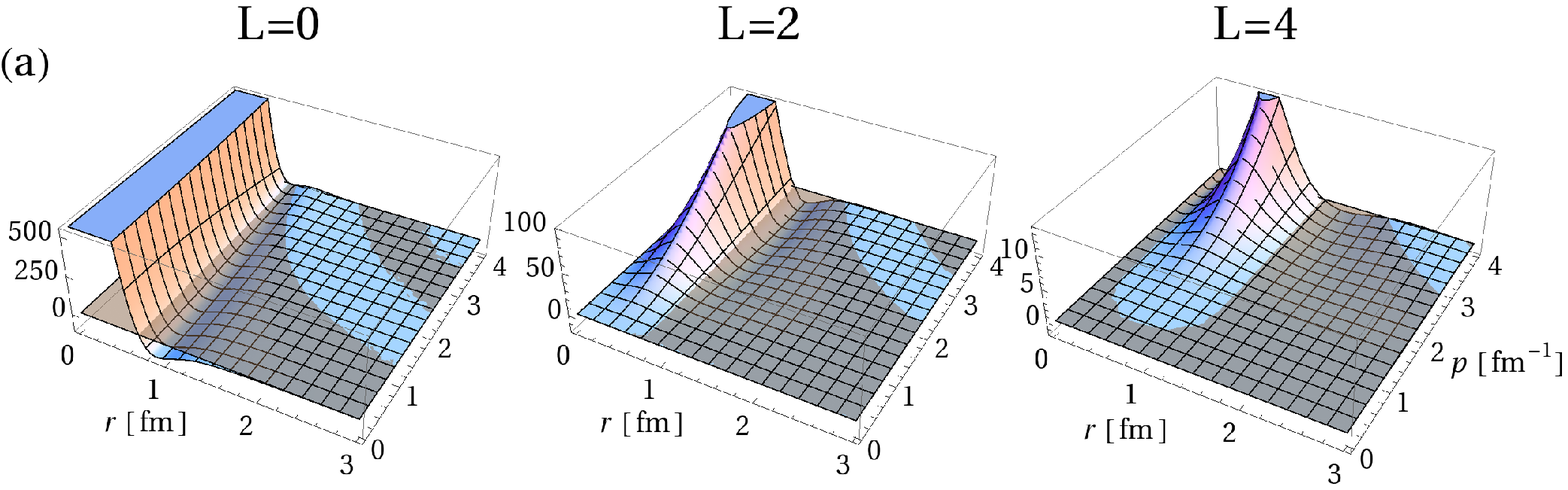}
\end{minipage}
\begin{minipage}[c]{0.9\textwidth}
\vspace{-6ex}
\includegraphics[width=0.9\textwidth]{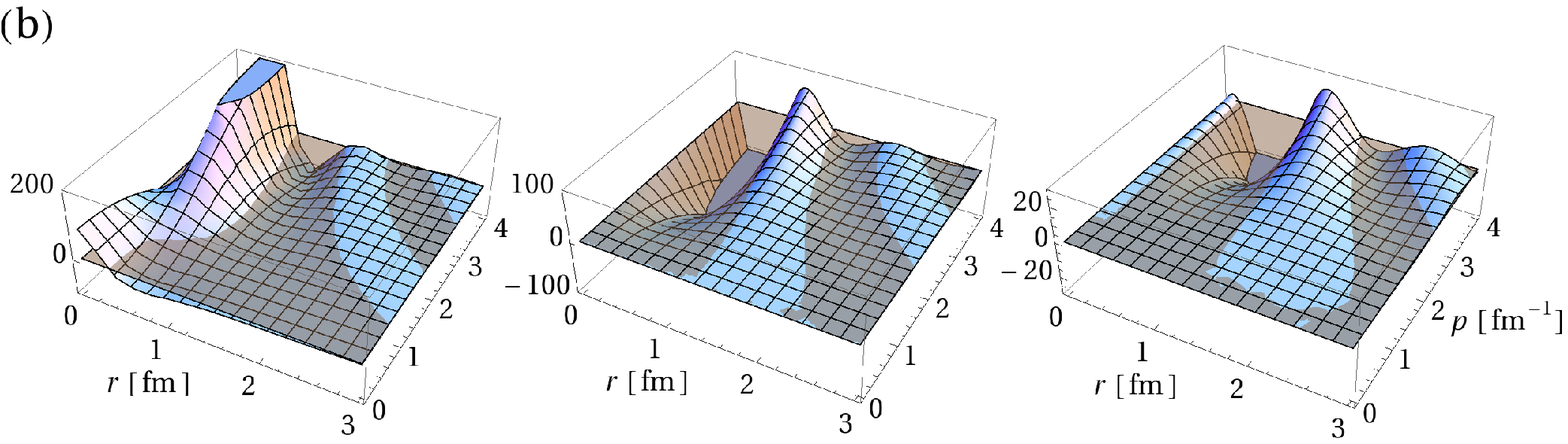}
\end{minipage}
\begin{minipage}[c]{0.9\textwidth}
\vspace{-6ex}
\includegraphics[width=0.9\textwidth]{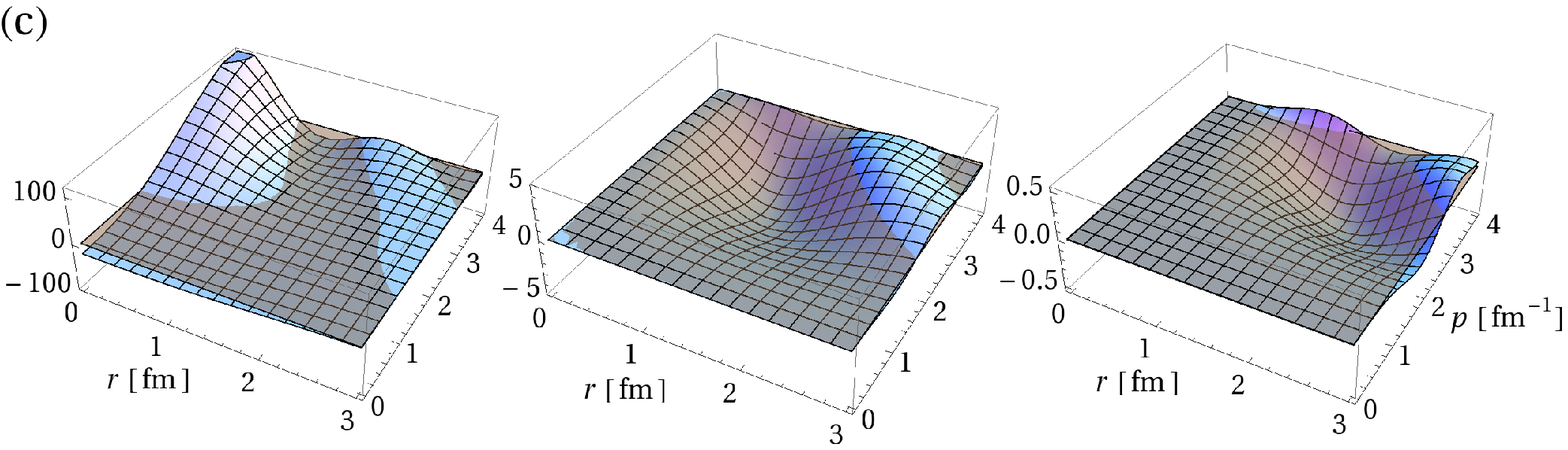}
\end{minipage}
\caption{\label{fig:Argonnerp}(Color online) Matrix elements $\braa{r;L,1}\op{V}\keta{p;L,1}$ (in MeV)
for (a) the bare Argonne potential, (b) the UCOM transformed Argonne potential, and (c)  the SRG 
transformed Argonne potential with a flow parameter of $\alpha=0.04\,\mathrm{fm}^4$ for $L=0,\,2,\,4$.}
\end{figure*}

In 2007 the similarity renormalization group method SRG \cite{bogner07b} was introduced into nuclear physics. This is 
another unitary transformation that provides a softened effective interaction while leaving the phase shifts  
unchanged. In case of SRG the interaction is evolved by means of a flow equation 
\begin{equation}\label{eq:SRG}
\frac{{\rm d} \op{V}_\alpha}{{\rm d}\alpha}=
\Big[\, \Big[\op{T},\op{V}_\alpha\Big],\op{H}_\alpha\Big]\ , \ \op{H}_\alpha=\op{T}+\op{V}_\alpha
\end{equation}
where with growing flow parameter 
$\alpha$ the matrix elements of the effective interaction $\op{V}_\alpha$ calculated in the eigenbasis
of the relative kinetic energy $\op{T}$ are driven towards a diagonal form. This implies a decoupling of 
low and high momenta, which is desirable for applications in low momentum Hilbert spaces.
For $S\se 0$ the Hamiltonian does not connect partial waves with 
different $L$ and hence the evolution proceeds in each angular momentum channel independently 
without coupling to other $L$. For $S\se 1$ the tensor interaction connects at most $L$ and $L+2$.
An advantage of SRG is that one can include many-body forces in a straightforward way.
As the evolution Eq.~\eqref{eq:SRG} is easily solved numerically on a momentum grid or in a 
basis representation,  SRG enjoys widespread use.  

In the following subsection we will first look at the $r,p$ matrix elements in 
angular momentum channels for different $L$. This is the analogue to $k,k'$ matrix
elements, which are shown in many publications, but here information on r-space is already visible.
They will also help in understanding the problems originating from treating the different
$L$ channels independently.

In Sec.~\ref{sec:rp-Lambda} we will combine the matrix elements from the $L$-channels to come finally to the 
phase-space representation of the potential as function of $r,p$ and 
powers of $\hat{r}\ndot\hat{p}$.

\subsection{$\textit{\textbf{r,p}}$ representation in partial waves $\textit{\textbf{L}}$}
\label{sec:rp-L}

We obtain the $r,p$ matrix elements $\braa{r;L,T}\op{V}\keta{p;L,T}$
by evaluating Eq.~\eqref{eq:rp-antisym} on an equidistant momentum grid with 
$\delta=0.1\,\mathrm{fm}^{-1}$ and $k_{max}=20\,\mathrm{fm}^{-1}$ for the 
Argonne potential. For the chiral N3LO potential and its SRG transformed effective versions  
$k_{max}=10\,\mathrm{fm}^{-1}$ is sufficient. 

In Fig.~\ref{fig:Argonnerp} the $r,p$ matrix elements of the Argonne potential and its 
UCOM and SRG transformed offspring are displayed for $L=0,2,4$. With larger $L$ they decrease in size
and tend to zero for $L > 8$. Both, the UCOM (Fig.~\ref{fig:Argonnerp}~(b)) and 
the SRG (Fig.~\ref{fig:Argonnerp}~(c)) transformation reduce the absolute magnitude of the
original Argonne matrix elements (Fig.~\ref{fig:Argonnerp}~(a)) substantially.

In order to compare both transformations quantitatively we choose the UCOM correlation
function $R_+(r)$  such that the transformed Argonne scattering solution, 
$\op{C}_r^{-1}\keta{\widehat{k;L,T}}$, for $k=0$ (zero energy) and $L=0$
equals the one obtained by the corresponding SRG transformation for $\alpha=0.04~{\rm fm}^4$.
Despite this adjustment, Fig.~\ref{fig:Argonnerp} shows that the UCOM transformation produces more 
momentum dependence and provides less softening. 

The so called natural extension to a local projection introduced in Ref.~\cite{wendt12} 
is for $L=0$ the same as the $r,p$ matrix element taken at $p=0$, 
which in turn is for $S\se 0$ the same as the genuine local projection Eq.~\eqref{eq:localproj}
because $\braa{r;L,T}\op{V}\keta{p\se 0;L,T}=0$ for $L\ne 0$. This fact has not been mentioned
explicitly in Ref.~\cite{wendt12}.

The behavior of the local projection of the SRG transformed 
matrix elements as a function of $\lambda=\alpha^{-1/4}$ is displayed in Fig.~2 of 
Ref.~\cite{wendt12} ($L\se 0$). In Fig.~\ref{fig:Argonnerp}~(c) the case 
$\alpha=0.04~{\rm fm}^4$ or $\lambda=2.24~{\rm fm}^{-1}$ is shown but now including the
$p$ dependence. One can see that the  pseudo local part $(p=0)$ drops fast with
increasing $L$ but for $L=0$ there is a strong quadratic momentum dependence at short distances
and the matrix element changes from negative to positive around the Fermi momentum.
In the following section \ref{sec:rp-Lambda} we discuss further the  local projection
given in Eq.~\eqref{eq:localproj}.

Another aspect that will be of importance for the phase-space representation, as discussed in the 
following section, is the occurrence of oscillations at large $rp$. 
This is due to the fact that the matrix element projects on a given $L$.
As can be seen from the definition (Eq.~\eqref{eq:rp-antisym}) even a constant potential proportional 
to the unit operator $\op{I}$, which does not create any force, exhibits these oscillations. 
(\,Eq.~\eqref{eq:rp-antisym} simplifies to $\braa{r;L,T}\op{I}\keta{p;L,T}=j_{L}(pr)$.\,)
The same holds true for example for the kinetic energy $\op{T}$. However after adding
up the $L$ contributions, as done in the phase-space representation,
one returns to $\vek{r}$ and $\vek{p}$,  and retrieves $\ps{I}(\vek{r},\vek{p})\se 1$ and  
$\ps{T}(\vek{r},\vek{p})\se\vek{p}^2/(2\mu)$, respectively.
Only in the phase-space representation one can judge forces 
($-\partial \ps{V}(\vek{r},\vek{p})/\partial\vek{r})$ or
contributions to the velocity ($\partial \ps{V}(\vek{r},\vek{p})/\partial\vek{p})$ properly.
Looking at individual $L$-channels can be misleading.

As has been realized already in the early days of realistic potentials \cite{reid68} 
different local potentials for each $L$ are helpful for fitting phase shifts but imply a 
complicated momentum dependence
because in this situation the spherical Bessel functions, which at large arguments fall 
off only like $1/(rp)$,
do not add up in a coherent way to yield smooth potentials. 

We dwell on this fact because in the following section, where the phase-space representation is
shown, we encounter these oscillations again. There they are  an undesired feature,
because they correspond to oscillating forces at $rp > 3$.

For $T\!=\!0$, $S\!=\!0$ only odd $L$ are allowed and the interaction is essentially repulsive.
This is also reflected in the UCOM and SRG transformed matrix elements not shown here.

\begin{figure*}[ht!]
\centering
\begin{minipage}[c]{\textwidth}
\includegraphics[width=0.9\textwidth]{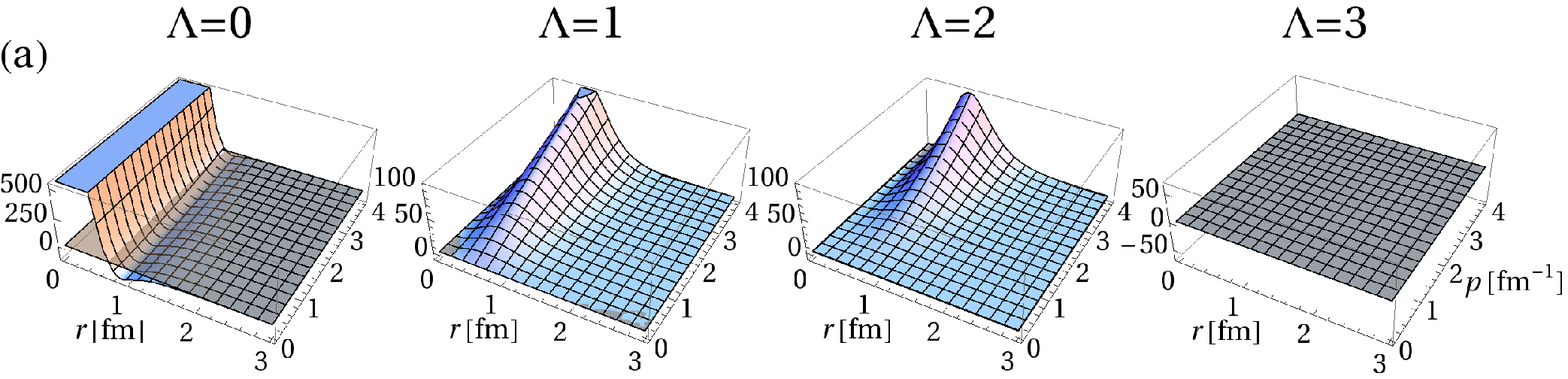}
\end{minipage}
\begin{minipage}[c]{\textwidth}
\vspace{-6ex}
\includegraphics[width=0.9\textwidth]{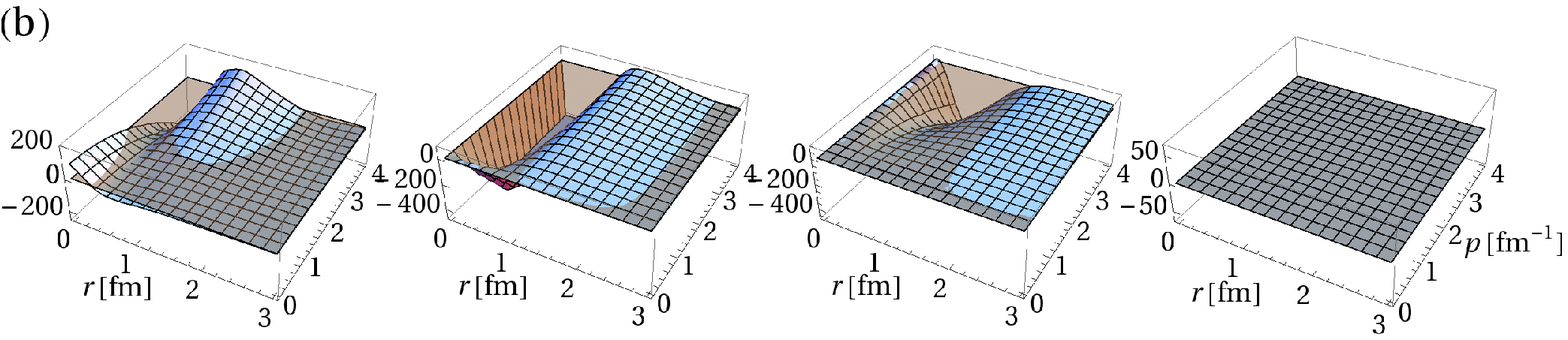}
\end{minipage}
\begin{minipage}[c]{\textwidth}
\vspace{-6ex}
\includegraphics[width=0.9\textwidth]{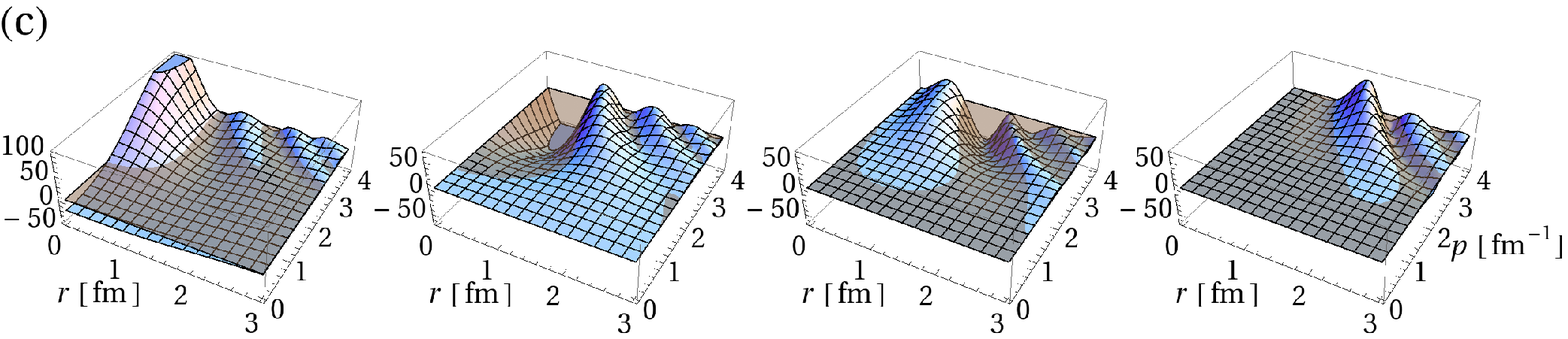}
\end{minipage}
\begin{minipage}[c]{\textwidth}
\vspace{-6ex}
\includegraphics[width=0.9\textwidth]{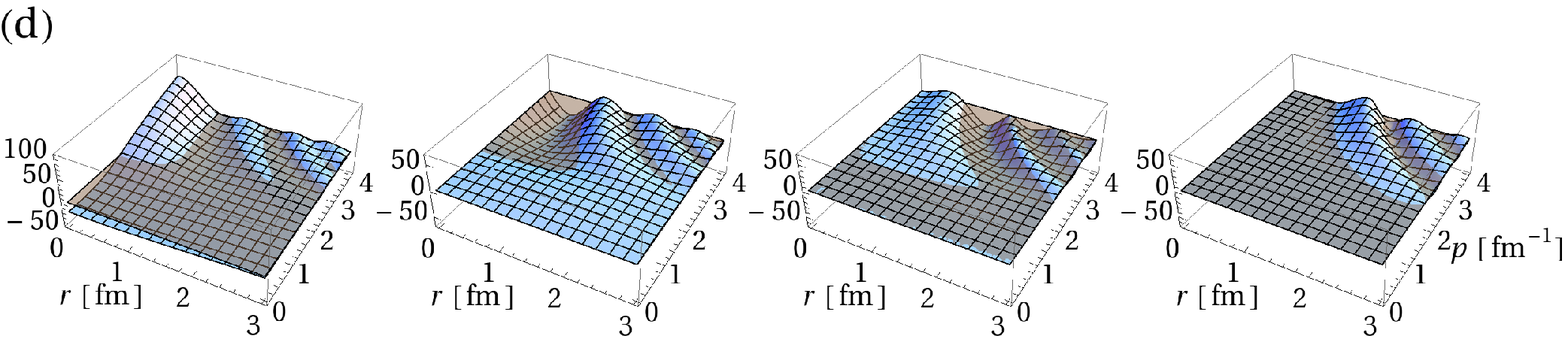}
\end{minipage}
\caption{\label{fig:Argonne}(Color online) Phase-space representation 
$\ps{V}^{T=1,\,\Lambda}(r,p)$ (in MeV) for (a) the bare Argonne potential, 
(b) the UCOM transformed Argonne potential, 
(c) the SRG transformed Argonne potential with a flow parameter of $\alpha=0.04\,\mathrm{fm}^4$, and 
(d)  the SRG transformed Argonne potential with a flow parameter of $\alpha=0.2\,\mathrm{fm}^4$.}
\end{figure*}

\begin{figure*}[ht!]
\centering
\begin{minipage}[c]{\textwidth}
\includegraphics[width=0.9\textwidth]{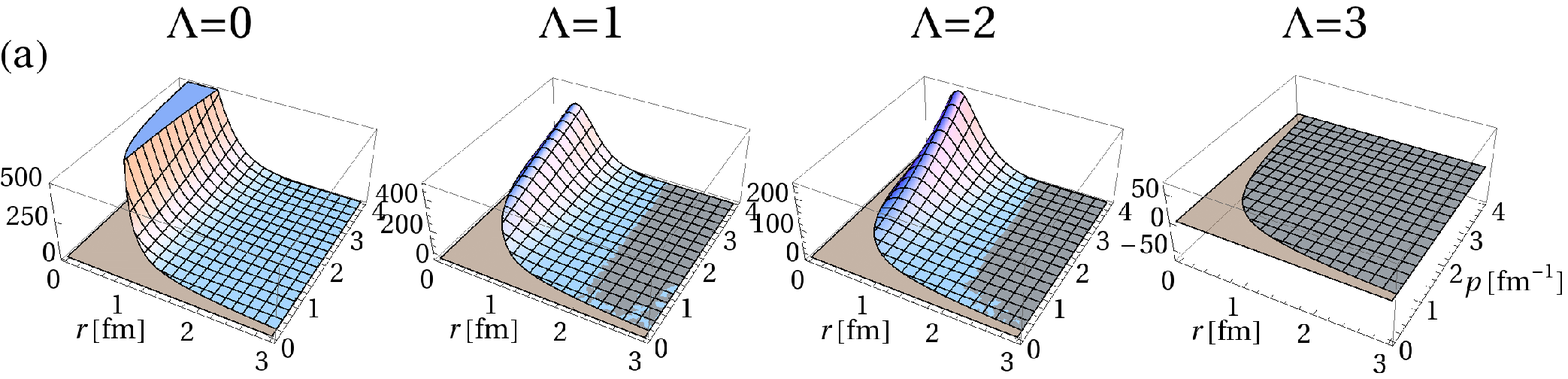}
\end{minipage}
\begin{minipage}[c]{\textwidth}
\vspace{-6ex}
\includegraphics[width=0.9\textwidth]{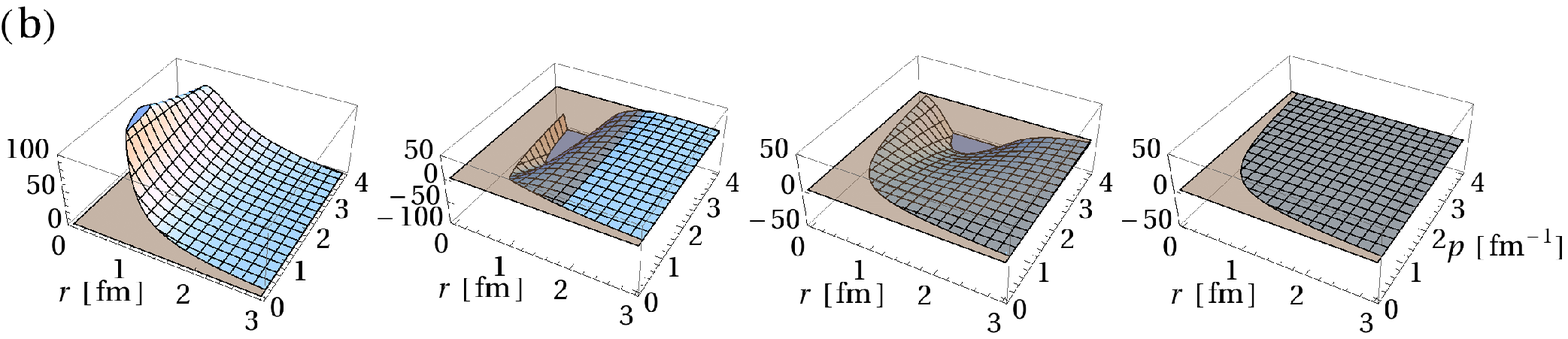}
\end{minipage}
\begin{minipage}[c]{\textwidth}
\vspace{-6ex}
\includegraphics[width=0.9\textwidth]{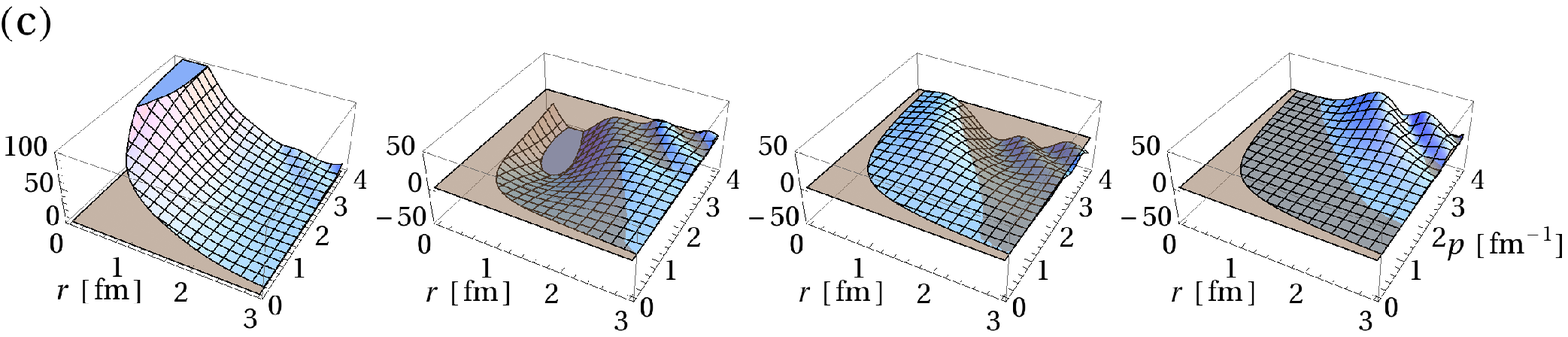}
\end{minipage}
\begin{minipage}[c]{\textwidth}
\vspace{-6ex}
\includegraphics[width=0.9\textwidth]{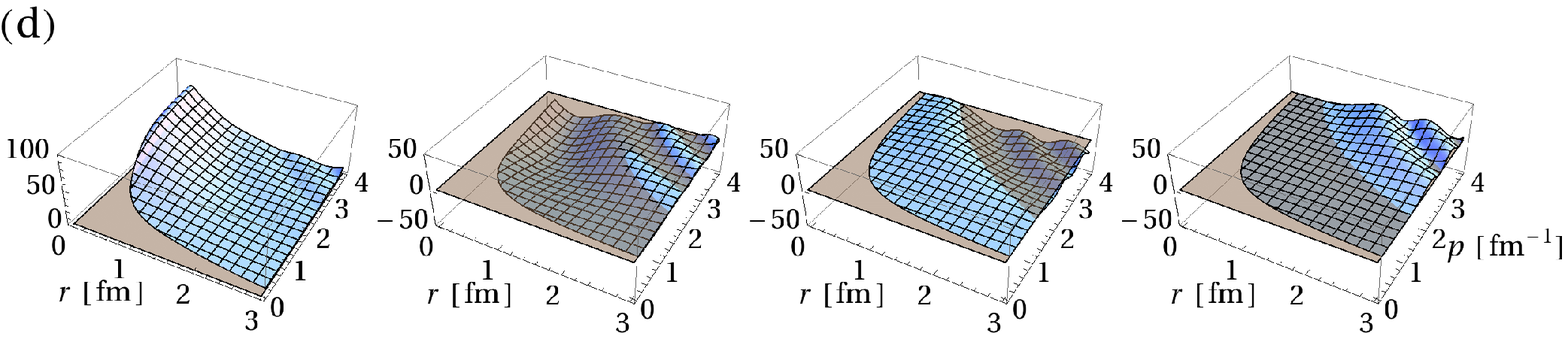}
\end{minipage}
\caption{\label{fig:ArgonneT0}(Color online) Phase-space representation 
$\ps{V}^{T=0,\,\Lambda}(r,p)$ (in MeV) for (a) the bare Argonne potential, 
(b) the UCOM transformed Argonne potential,
(c) the SRG transformed Argonne potential with a flow parameter of $\alpha=0.04\,\mathrm{fm}^4$, and 
(d)  the SRG transformed Argonne potential with a flow parameter of $\alpha=0.2\,\mathrm{fm}^4$.}
\end{figure*}

\begin{figure*}[ht!]
\centering
\begin{minipage}[c]{\textwidth}
\includegraphics[width=0.9\textwidth]{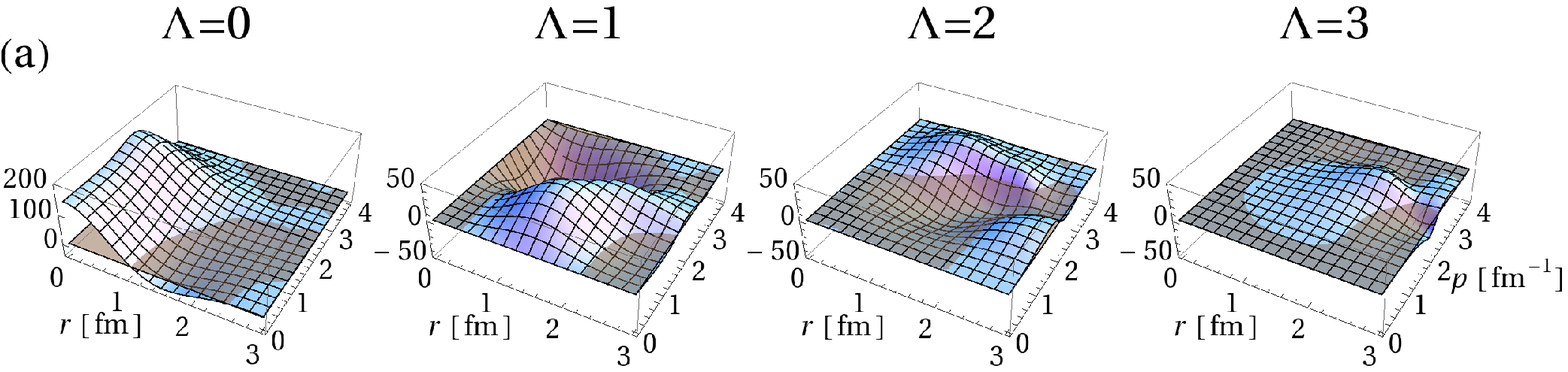}
\end{minipage}
\begin{minipage}[c]{\textwidth}
\vspace{-6ex}
\includegraphics[width=0.9\textwidth]{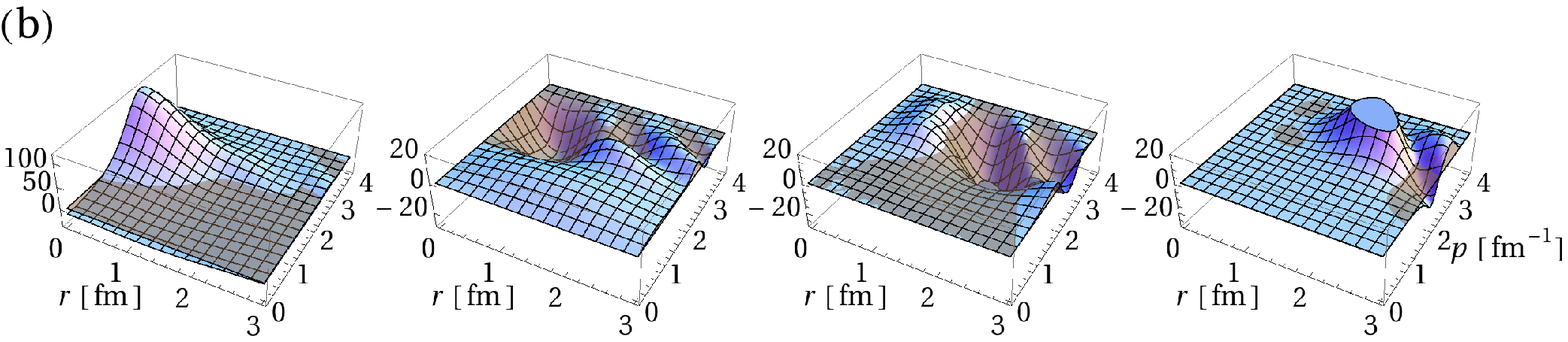}
\end{minipage}
\begin{minipage}[c]{\textwidth}
\vspace{-6ex}
\includegraphics[width=0.9\textwidth]{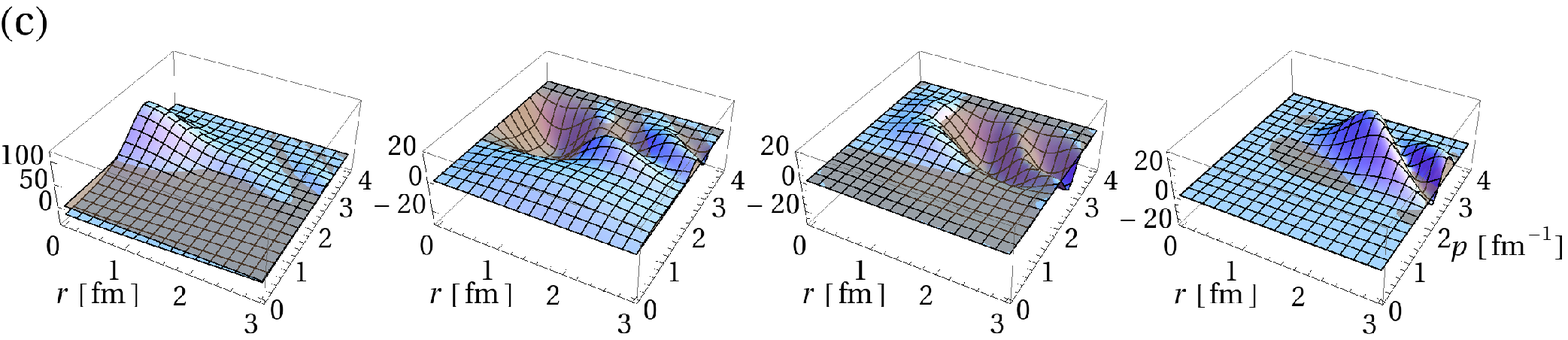}
\end{minipage}
\caption{\label{fig:n3lo}(Color online) Phase-space representation $\ps{V}^{T=1,\,\Lambda}(r,p)$ (in MeV) 
for (a) the bare chiral N3LO potential, (b) the SRG transformed potential with a flow parameter of 
$\alpha=0.04\,\mathrm{fm}^4$, and (c) $\alpha=0.2\,\mathrm{fm}^4$.}
\end{figure*}

\subsection{Phase-space representation in terms of Legendre polynomials or powers of 
$\hat{\op{r}}\ndot\hat{\op{p}}$}
\label{sec:rp-Lambda}

As the $r,p$ partial wave matrix elements become negligibly small for large $L$ we only include contributions
to the component $\ps{\mathcal{V}}^{T,\Lambda'}\!(r,p)$ of the antisymmetrized phase-space representation 
in Eq.~\eqref{eq:VLambda-rp-antisym} up to $L=8$, and hence $L'$ goes up to
$\Lambda'+8$. 

We start with the bare Argonne potential, which for $S\!=\!0$ consists of a local central part and a 
quadratic angular momentum term. As discussed in Sec.~\ref{sec:antisymmPSR}, one needs in this case 
$\Lambda_V=2$ in the expansion of $\ps{V}^T(\vek{r},\vek{p})$ to obtain a complete description of 
$\ps{\mathcal{V}}^T(\vek{r},\vek{p})$. 
The phase-space representation of the bare $S\!=\!0,\,T\!=\!1$ Argonne potential is presented in 
Fig.~\ref{fig:Argonne}~(a). The $\Lambda=0$ part is dominated by the local strong short-range repulsion
of the potential, so that the $p$-dependence originating from the $\op{\vek{L}}\phantom{}^2$ term is barely
visible. 
The $\Lambda=1$ and $2$ parts contain only contributions from the $\op{\vek{L}}\phantom{}^2$ and one observes the 
characteristic linear $p$-dependence for $\Lambda=1$ and quadratic $p$-dependence for $\Lambda=2$ 
(see Eq.~\eqref{eq:partL2}). 

Next, we investigate in Fig.~\ref{fig:Argonne}~(b) the phase-space representation of the UCOM 
transformed Argonne potential. It contains besides the local and the quadratic angular momentum term 
also a term quadratic in momentum:
\begin{eqnarray*}
\op{V}^{\mathrm{UCOM}}_{S=0,T}&=&V_{0,T}^C(\op{r})+V^{L2}_{0,T}(\op{r})\,\vek{\op{L}}^{2}\\
&&+\frac{1}{2}\left(\vek{\op{p}}^{2}V^{p2}_{0,T}(\op{r})+V^{p2}_{0,T}(\op{r})\,\vek{\op{p}}^{2}\right).
\end{eqnarray*} 
In this case the expansion is also limited to $\Lambda_V\!=\!2$. 
While the $\Lambda\!=\!2$ part contains only contributions from 
the angular momentum term, $\Lambda\!=\!1$ has now also input from the quadratic momentum term. 

In contrast to the bare Argonne potential, the $\Lambda\!=\!0$ part of the UCOM transformed Argonne potential 
shows for $p\!=\!0$ a much softer potential, which was of course the intention of UCOM. 
On the other hand a clear $p$-dependence originating 
from the new $\vek{\op{p}}^{2}$ term of the UCOM transformed potential appears. 
One might wonder about the behavior of the phase-space representation 
at small distances $r$ and high momenta $p$ where it becomes strongly negative. 
But one has to keep in mind that the $\vek{\op{p}}^{2}$ term in the UCOM potential 
originates from the transformation of the kinetic energy $(\op{C}_r^{-1}\op{T}\,\op{C}_r - \op{T}\,)/(2\mu)$.
The full Hamilton function, obtained by adding the kinetic energy 
$\ps{T}(\vec{r},\vec{p})\!=\!p^2/(2\mu)$ again, is still repulsive in this region.
In a classical view the momentum dependencies change the relative velocity of the particle pair. 
In the repulsive core area they speed up and spend less time there so that the density
is depleted, which corresponds to the correlation hole in the wave function.  

The phase-space representation therefore shows in a very lucid way how the unitary UCOM transformation,
which does not alter the phase shifts, succeeds in softening the potential by introducing 
momentum dependencies. 

The phase-space representation of $\op{V}_{\rm UCOM}$ in Fig.~\ref{fig:Argonne}~(b) was 
obtained by first calculating its antisymmetrized matrix elements
$\braa{k;L,T}\op{V}_{\rm UCOM}\keta{k';L,T}$ and then performing the procedure explained in
Sec.~\ref{sec:antisymmPSR} to extract the phase-space representation of  $\op{V}_{\rm UCOM}$ 
from its matrix elements.
The direct way to get $\ps{V}(\vek{r},\vek{p})$ from the known operator form of $\op{V}_{\rm UCOM}$
was used to test if the procedure is working properly.

In Fig.~\ref{fig:Argonne}~(c) and (d) we display the phase-space representation of the SRG transformed 
Argonne potential for two values of the flow parameter. 
For $\alpha=0.04\,\mathrm{fm}^4$, a typical value used in many-body calculations, the SRG transformed interaction
looks quite different from the UCOM one, although the correlation function
$R_+(r)$, which determines the UCOM transformation, was taken to map the zero energy scattering solution 
for $L=0$ onto the SRG solution for $\alpha=0.04\,\mathrm{fm}^4$ 
(as explained in Sec.~\ref{sec:rp-L}). 

\begin{figure*}[ht!]
\centering
\begin{minipage}[c]{0.45\textwidth}
\includegraphics[width=0.99\textwidth]{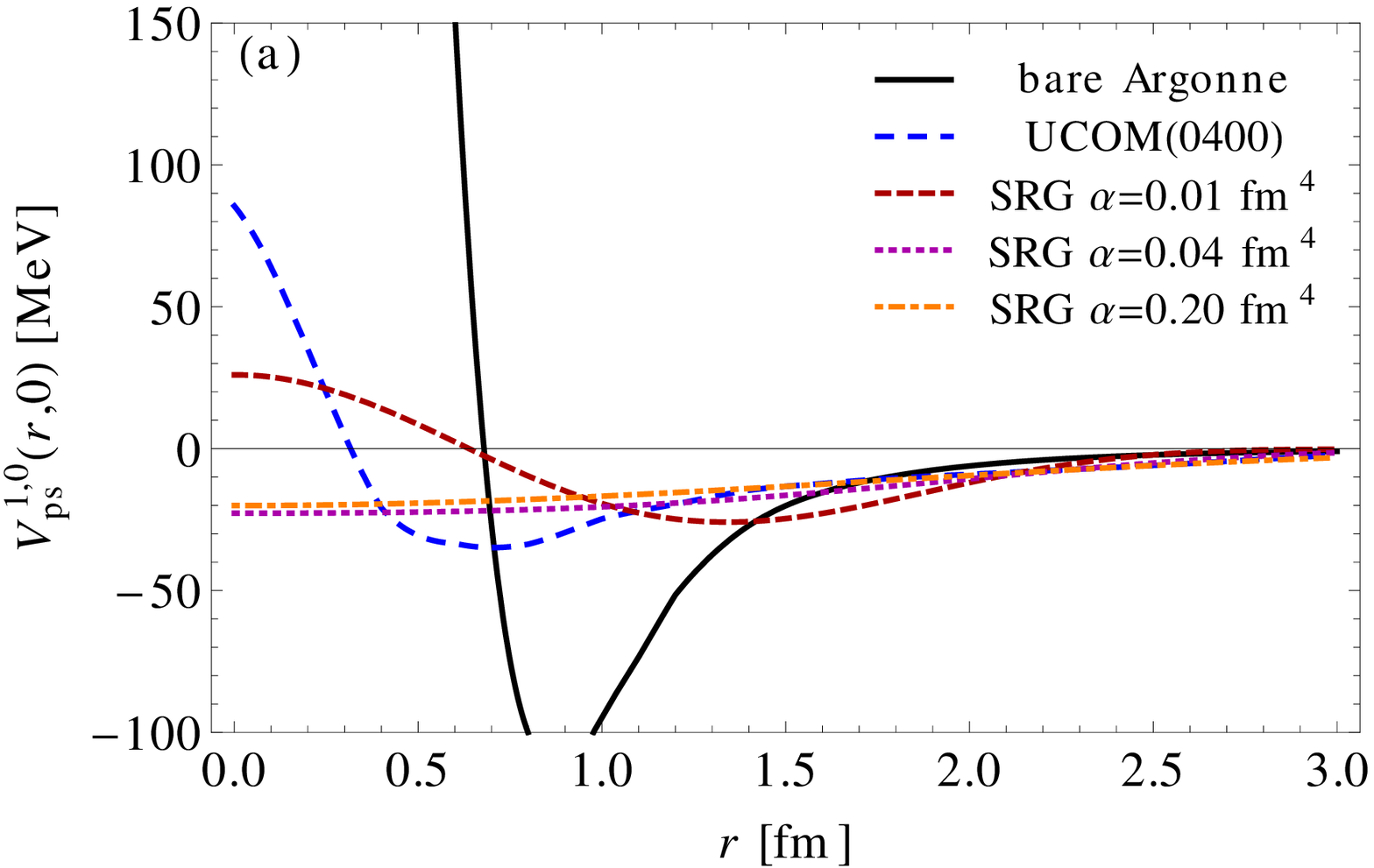}
\end{minipage}
\begin{minipage}[c]{0.45\textwidth}
\includegraphics[width=0.99\textwidth]{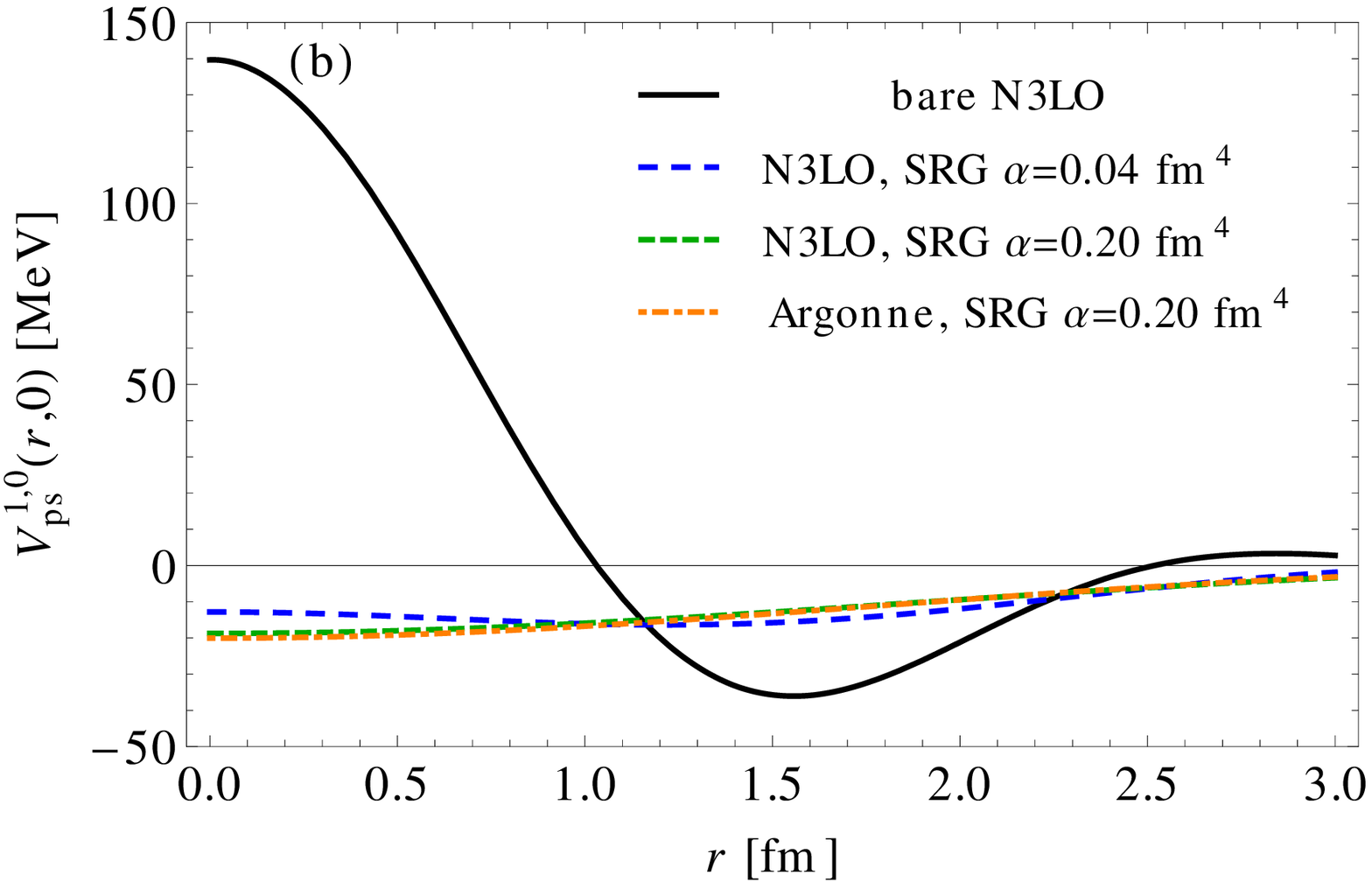}
\end{minipage}
\caption{\label{fig:core}(Color online) Phase-space representation for $p=0$, 
$\ps{V}^{T=1,\,\Lambda=0}(r,0)$ (in MeV): 
(a) The Argonne potential and various effective potentials based on it, 
(b) the N3LO potential and its SRG transformed version with flow parameter 
$\alpha=0.04\,\mathrm{fm}^4$ and $\alpha=0.2\,\mathrm{fm}^4$ compared with the SRG 
transformed Argonne potential with $\alpha=0.2\,\mathrm{fm}^4$.}
\end{figure*}

The first striking difference is that for $\Lambda=0$
and $p=0$ the SRG potential is at all distances $r$ attractive while the UCOM potential is still repulsive
for $r<0.7\ {\rm fm}$. The original Argonne Potential turns repulsive at $r=0.9\ {\rm fm}$.
The second conspicuous feature of the SRG potential is its long range, 
much longer than that of the original potential. This aspect and its consequences will be
discussed further in Fig.~\ref{fig:core}.
  
With increasing momentum $p$ the $\Lambda=0$ SRG transformed potential becomes more and more 
repulsive at short distances $r$. 
A further interesting, but annoying, result is that for all $\Lambda$ 
at large $rp$ oscillations develop, quite at variance with the UCOM transformed interaction.
SRG evolves the effective interaction in the different $L$-channels independently, while UCOM uses operators.
As already indicated in the previous Sec.~\ref{sec:rp-L} independent treatment of the 
$L$-channels may lead to complicated momentum dependencies showing up at larger values of $rp$.
The speed of the evolution of $\op{V}_\alpha$ may very well differ for different $L$ because
it is determined by the strength of the generator $[\op{T},\op{V}_\alpha]$ which increases
with the curvature of $\op{V}_\alpha$. Or in other words, the larger an off-diagonal matrix element
$\braa{k;L,T}\op{V}_\alpha\keta{k';L,T}$ is, the faster it decreases in size. 
Therefore $L\!=\!0$ is smoothened differently from $L\!=\!2$ and both channels cannot be interpreted
anymore as being projected from a potential with a simple momentum dependence.

In $A$-body space the SRG transformation leaves the observables invariant if the evolution
Eq.~\eqref{eq:SRG} includes up to $A$-body operators. That means that the oscillations at 
large $rp$ in 2-body space have to be compensated by induced many-body forces.
This might explain the fact, that large scale shell model calculations 
for larger mass numbers show significant contributions from 3-body and higher terms in the SRG evolved effective interaction,
see Ref.~\cite{roth13}.

For the large flow parameter $\alpha=0.2\,\mathrm{fm}^4$, corresponding to a very soft potential,
the oscillations are still present, see Fig.~\ref{fig:Argonne}~(d). 


Fig.~\ref{fig:ArgonneT0} displays the same phase-space representations as Fig.~\ref{fig:Argonne}
but for $T\!=\!0$, i.e. the UCOM transformed one and the SRG descendants for 
$\alpha=0.04\,\mathrm{fm}^4$ and $\alpha=0.2\,\mathrm{fm}^4$.
Both, the UCOM and SRG transformation soften the strong
repulsion but SRG produces again oscillations.
All considerations and discussions for the even channels also apply here for the odd channels
and need not be repeated.

The phase-space representation for $T\se 0$, $S\se 0$ is not shown for $rp<0.5$ because the spatial part 
of the wave function has to be antisymmetric and thus only odd $L$ contribute. As there is 
no information for $L\se 0$ available, our method was not able
to extract numerically stable results from the matrix elements in the region $rp<0.5$.
But this is also not necessary because states with $L \ge 1$ can not probe this area.

The last example shows the phase-space representation of the 
chiral N3LO potential for $T\!=\!1$, $S\!=\!0$ from Entem and Machleidt 
\cite{entem03} (Fig.~\ref{fig:n3lo}~(a)).  %
Its two-body part includes pion-exchange contributions and local and non-local contact terms.
Furthermore, the potential is regularized by a (smooth) momentum cutoff at $500\,\mathrm{MeV/c}$. 
To describe the momentum dependence of the N3LO potential, we need terms at least up to $\Lambda_V=3$ 
in the expansion of phase-space representation. 
For $\Lambda=0$ and $p=0$ the phase-space representation shows repulsion at short distances, although much weaker than in case of the Argonne interaction, and attraction at larger distances.
With increasing $p$ the phase-space representation clearly visualizes the momentum cutoff around $p=2.5\,\mathrm{fm}^{-1}$. 
For higher $\Lambda$ one sees momentum structures that are more complicated than those of the 
Argonne potential.

Fig.~\ref{fig:n3lo}~(b) and (c) display the phase-space representations of the SRG transformed
N3LO potential for $\alpha=0.04\,\mathrm{fm}^4$ and $\alpha=0.2\,\mathrm{fm}^4$. 
Below $p\approx 1.5\,\mathrm{fm}^{-1}$ SRG clearly softens the bare N3LO and 
the transformed potentials are rather smooth.
But for larger momenta $p$ they develop a lively $r$ and $p$ dependence, that is even more pronounced 
than for the bare N3LO interaction. 
The UCOM transformation of the N3LO interaction is shown because it has not been investigated yet.

Finally we show in Fig.~\ref{fig:core} the local projections of the potentials as defined in 
Eq.~\eqref{eq:localproj}. The results agree with those in 
Fig.~2 of Ref.~\cite{wendt12} as explained above.
The bare N3LO potential is much less repulsive than the bare Argonne potential but a repulsive force, as given by the gradient of the potential,
sets in already for $r<1.55\,\mathrm{fm}$. Furthermore there is an overshooting above
$r=2.5\,\mathrm{fm}$ despite the fact that also in N3LO the long range part is 
determined by pion exchange. This overshooting is a fringe effect, which results from
the cutoff in momentum space. Due to the power counting in momentum one feels obliged
to sharpen the cutoff function $\exp\{-(p/\Lambda_{\rm cutoff})^{2n}\}$ with higher
order $n$. But this leads to more pronounced fringes, which we regard as problematic. 

Ref.~\cite{wendt12} emphasizes the universal behavior of the SRG evolved 
interactions seen in the similarity of the local projections of the evolved Argonne and N3LO potentials. 
By comparing Fig.~\ref{fig:Argonne}~(d) with Fig.~\ref{fig:n3lo}~(c) one realizes
that the universality is not only seen at $p\se 0$ but extends to about
$p\approx 1.2\,\mathrm{fm}^{-1}$. One could not anticipate this by looking at 
the pseudo-local projection 
for $L\se 2$ displayed in Fig.~4 of Ref.~\cite{wendt12},
where the Argonne and the N3LO case still differ substantially.

\subsection{On local potentials} 
\label{sec:localpot}

There are infinitely many non-local potentials with identical phase shifts, which differ 
in their degree of non-locality or momentum dependence. They also differ in the size of 
their corresponding three- and more-body interactions that are needed to correctly describe many-body systems.
In such a situation the question arises how to constrain the form of the nuclear interaction
beyond general symmetry considerations such that it is useful and numerically manageable
for a wide variety of low energy nuclear structure applications. 

Our intuition favors local potentials because in daily life they are very successful for
describing the forces between two objects. The reason is that usually the relative velocity $v$
of the two objects is small. There are essentially two smallness conditions, the first is that $v$ is
small compared to the speed of light $c$, so that retardation effects can be neglected and 
an instantaneous action at a distance picture with a static potential is suited, 
like the Coulomb potential.
The lowest order relativistic correction, proportional to $v/c$, is the Lorentz force, or for 
particles with spin and magnetic moment, like nucleons or electrons, the spin-orbit interaction. 
The next order $(v/c)^2$ terms originate from relativistic kinematics or 
the relativistic energy-momentum relation.
In the case of nucleons a typical speed is 
given by the Fermi momentum of 260~MeV/c which corresponds to $v/c\approx 0.3$. Therefore
relativistic effects are expected to be small but not negligible.

The second smallness condition comes when one regards the interaction between two objects that
are composed of particles bound together by their mutual microscopic interaction, like
two atoms or two nucleons. The potential between the electrically or color neutral particles is then
of the van der Waals type and originates from the microscopic electric or microscopic strong
interaction between the constituents. In that case one can imagine for the pair at rest a 
static local potential as function of the distance $r$ between their center of masses. 
The local potential is defined as the total energy of the two-body system minus the total energy
at large distance, where the two objects do not interact anymore. 
This is a useful picture if the relative velocity is much smaller
than the intrinsic velocities of the constituents (the electrons in case of atom-atom potentials)
so that the mutual polarization can readjust instantaneously.
In the case of atoms the chemical reactions proceed at small enough velocities so that one
can usually work with local potentials. 

For nucleons the situation is different. If one takes as an estimate of the time scale for polarization 
the excitation energy of the $\Delta$-resonance of 300~MeV, which corresponds to an oscillation time
of 4~fm/$c$, one would argue that for a static local potential a typical interaction time 
or passage should be much longer than that. 
For traveling a distance of 2~fm at Fermi velocity one needs about 7~fm/$c$,
which is of the same order as the oscillation time and the polarization cannot adjust instantaneously.
This simple estimate indicates already that virtual excitations of resonances should play
an important role and lead to non-localities or momentum dependencies and many-body interactions when
the resonances are not treated as explicit degrees of freedom.

From these simple arguments it is not surprising that a purely local potential cannot fit
the phase shifts of nucleon-nucleon scattering up to $E_{\rm LAB}=300$~MeV or the corresponding
$v/c\approx 0.5$. But it still makes sense to expand in $v/c$ or $p/m$ and keep terms up to
$p^2$ or $p^4$ if needed. 

Reproducing measured phase shifts is only one part of the task in modeling the nuclear
interaction. In an interacting many-body system like a nucleus, the nucleon pairs are not on-shell,
meaning that they neither have a sharp energy nor is their
relative momentum in the same way related to their mutual energy as in the scattering process. 
Furthermore, as they are interacting simultaneously with other nucleons their relative 
angular momentum is also not in an eigenstate of $\op{\vek{L}}\phantom{}^2$. 
Therefore information coming from bound many-body states can and should be used to
reduce the arbitrariness in non-local versus local parts or in off-shell properties.

For these reasons it makes sense to start with an operator structure that has local radial
dependencies augmented with low powers in $\op{\vek{p}}$ and $\op{\vek{L}}$ in accordance
with the global symmetries like invariance under parity, rotation and translation. 
The power expansion together with spin and isospin structures can then be used to
establish perturbative counting schemes based on chiral symmetry.
Keeping an operator based interaction of this type will prevent the introduction of
uncontrolled off-shell behavior in the many-body case and also hopefully avoids the need for
$n$-body interactions with ever increasing $n$.

It is very desirable to develop schemes that do not produce complicated non-localities
or momentum dependencies, neither when choosing the resolution scale nor when re-normalizing
to low momentum many-body Hilbert spaces. 
Recently there have been attempts \cite{gerzelis13,gerzelis14,lynn14} to use local regulators in
chiral effective field theories. The non-smooth behavior of the resulting potential \cite{gerzelis14}
might indicate deficiencies of the presently employed regulators.

\section{Summary}

Reproducing nuclear phase shifts up to $E_{\rm LAB}=300\,$MeV with a non-relativistic 
Schr\"odinger equation requires 
two-body interactions that are non-local in coordinate space or in other words momentum dependent. 
As a rule of thumb one can say that potentials with a weak momentum dependence (almost local) 
exhibit a strong short range repulsion while softer potentials have a stronger momentum
dependence. Although all realistic potentials, may they have stronger or weaker momentum dependence,
describe the nuclear two-body problem equally well, they can give quite different results in the
many-body case. This is to some extend compensated in the many-body system by appropriate many-body forces that come
along with the respective two-body interaction. 

While many-body Quantum Monte Carlo methods \cite{pieper05, pieper01mp} work best with local potentials, 
methods based on many-body Hilbert spaces spanned by Slater determinants converge better
when softer interactions are
used. The reason is that a Slater determinant basis cannot represent with sufficient accuracy the 
short range repulsive and tensor correlations induced by a hard interaction. 

As these correlations at nuclear densities are mainly of short ranged two-body type 
and universal in the sense that they do not depend on the mass number of the nucleus or its 
excitation energy \cite{feldmeier11,schiavilla07} it should
be possible to renormalize them into the effective interaction without strong induced 
three- or higher-body contributions.
Therefore a useful strategy would be to renormalize a given hard interaction only at small distances. 
Then it should be sufficient to include only the three-body part of the induced many-body interactions, 
as the probability to find four and more particles simultaneously close together is very small. 
This concept is pursued by UCOM but the calculation of induced many-body interactions in 
UCOM is very complicated. On the other hand a SRG transformation that explicitly takes 
the universality of short-range correlations into account has still to be developed. 

Another approach is to limit the resolution scale already in the derivation of the interaction and 
to seek a smallness parameter to order the possible contributions that are allowed by symmetries.
However, if one wants to retain the phase shifts the above rule of thumbs applies in any case.

In many of these attempts one represents the interaction in terms of momentum space matrix elements
because they are well suited for calculating phase shifts and many-body matrix elements. Also similarity 
renormalization group methods naturally work within such a basis.
A drawback in doing that is that the explicit operator structure of the interaction and the intuitive
picture is lost.

This paper attempts to recover the intuition by extracting the Kirkwood phase-space representation
of the interaction from a given set of matrix elements. With this phase-space representation one 
can visualize for example
in a quantitative way the difference in momentum dependence between the softer N3LO chiral potential
and the hard Argonne potential. One also sees how UCOM or SRG renormalization transforms
short range repulsion into momentum dependence (retaining the phase shifts).
One observes that the SRG transformation shows a universal behavior not only for the
local projection, i.e. at momentum $p=0$, but results in a dominant $\Lambda=0$ part,
which depends only on $r$ and $p$, with a weak quadratic $p$-dependence for momenta up to
$p\approx 1.2$\,fm$^{-1}$. 
The $\Lambda=1,2,3$ parts are an order of magnitude smaller or in other words the dependence on 
$\hat{r}\ndot\hat{p}$, $(\hat{r}\ndot\hat{p})^2$ and $(\hat{r}\ndot\hat{p})^3$ is weaker indicating 
that an ansatz with low powers in  $\op{\vek{L}}\phantom{}^2$ should be sufficient.

But the phase-space representation uncovers also peculiarities that are hidden in the matrix element 
representation, like the effects of inconsiderate ways of implementing momentum cutoffs or 
shortcomings of the renormalization. 
One sees for example that the SRG transformation develops an irregular phase-space behavior for 
momenta $p$ larger than about 1.5\,fm$^{-1}$ for the Argonne potential
and about 1.2\,fm$^{-1}$ for the N3LO potential, which limits the range of the smooth universal
behavior. One should investigate if an appropriate choice of the 
SRG generator can extend the range, because 1.2\,fm$^{-1}$ is still below the 
Fermi momentum $k_F=1.32\,\mathrm{fm}^{-1}$. 

In continuation of the work presented in this paper the phase-space representation will be extended 
to include spin-orbit and tensor forces in the $S=1$ channels.

\section{Acknowledegments}
This work was supported by the Helmholtz Alliance Program of the Helmholtz Association, contract 
HA216/EMMI "Extremes of Density and Temperature: Cosmic Matter in the Laboratory".\\

\appendix
\section{Partial wave decomposition of the phase-space representation\label{app:pw}}
In the following we calculate $\e^{-2\i\vek{r}\cdot\vek{p}}\ps{V}^T(\vek{r},-\vek{p})$ by
expanding first $\e^{-2\i\vek{r}\cdot\vek{p}}$ and then $\ps{V}^T(\vek{r},-\vek{p})$ in 
spherical harmonics. 
\begin{eqnarray}
\e^{-2\i\vek{r}\cdot\vek{p}}=
4\pi \sum_{L,M}(-\i)^L j_L(2rp) Y^L_M(\hat{p}){Y^L_M}^*\!(\hat{r})\, ,
\label{eq:2pw}
\end{eqnarray}
where we used 
\begin{align}\label{eq:minus}
Y^L_M(-\hat{p})=(-1)^L Y^L_M(\hat{p})\, .
\end{align}
The expansion of $\ps{V}^T(\vek{r},-\vek{p})$ in Legendre polynomials or spherical harmonics
is given by 
\begin{align}
\ps{V}^T&(\vek{r},-\vek{p})=
\sum_{\Lambda'} \ps{V}^{T,\Lambda'}(r,p)\ \i^{\Lambda'}P_{\Lambda'}(-\hat{r}\ndot\hat{p})\nonumber \\
&=4\pi\!\sum_{\Lambda',\mu'} \ps{V}^{T,\Lambda'}(r,p)\,\frac{(-\i)^{\Lambda'}}{2\Lambda'+1}
                               Y^{\Lambda'}_{\mu'}(\hat{p})\,{Y^{\Lambda'}_{\mu'}}^*\!(\hat{r})\, ,
\label{eq:VT} 
\end{align}
where the identity  Eq.~\eqref{eq:PlY} and Eq.~\eqref{eq:minus}) were employed. 

\begin{widetext}
The next step consists in combining the spherical harmonics
\begin{subequations}
\begin{eqnarray}
Y^{L}_{M}(\hat{p}) \, Y^{\Lambda'}_{\mu'}(\hat{p})&=&
\frac{1}{\sqrt{4\pi}}\sum_{\Lambda,\mu}
\sqrt{\frac{(2L+1)(2\Lambda'+1)}{(2\Lambda+1)}}
\ClebschGordan{L}{M}{\Lambda'}{\mu'}{\Lambda}{\mu}
\ClebschGordan{L}{0}{\Lambda'}{0}{\Lambda}{0}
{Y^{\Lambda}_{\mu}}(\hat{p})  \\
{Y^{L}_{M}}^*\!(\hat{r}) \, {Y^{\Lambda'}_{\mu'}}^*\!(\hat{r})&=&
\frac{1}{\sqrt{4\pi}}\sum_{\Lambda'',\mu''}
\sqrt{\frac{(2L+1)(2\Lambda'+1)}{(2\Lambda''+1)}}
\ClebschGordan{L}{M}{\Lambda'}{\mu'}{\Lambda''}{\mu''}
\ClebschGordan{L}{0}{\Lambda'}{0}{\Lambda''}{0}
{Y^{\Lambda''}_{\mu''}}^*\!(\hat{r})  \, .
\end{eqnarray}\label{eq:YYnachY*}
\end{subequations}
With these relations we obtain by multiplying Eq.~\eqref{eq:2pw} and  Eq.~\eqref{eq:VT}
\begin{align}
\e^{-2\i\vek{r}\cdot\vek{p}}\,\ps{V}^T(\vek{r},-\vek{p})=& 
4\pi \sum_{L, \Lambda'}\ps{V}^{T,\Lambda'}(r,p)\ (-\i)^{L+\Lambda'}j_L(2rp)
\sum_{\Lambda, \Lambda''}\sum_{\mu, \mu''}
\frac{(2L+1)}{\sqrt{(2\Lambda+1)(2\Lambda''+1)}}
\ClebschGordan{L}{0}{\Lambda'}{0}{\Lambda}{0}
\ClebschGordan{L}{0}{\Lambda'}{0}{\Lambda''}{0}
\times \nonumber \\
&{Y^{\Lambda''}_{\mu''}}^*\!(\hat{r})\,{Y^{\Lambda}_{\mu}}(\hat{p})\
\sum_{M,\mu'}
\ClebschGordan{L}{M}{\Lambda'}{\mu'}{\Lambda}{\mu}
\ClebschGordan{L}{M}{\Lambda'}{\mu'}{\Lambda''}{\mu''} \, .
\label{eq:exc1}
\end{align}
By using the orthogonality relation 
\begin{eqnarray}
\sum_{M,\mu'}\ClebschGordan{L}{M}{\Lambda'}{\mu'}{\Lambda}{\mu}
\ClebschGordan{L}{M}{\Lambda'}{\mu'}{\Lambda''}{\mu''}=\delta_{\Lambda\Lambda''}\delta_{\mu\mu''}
\end{eqnarray}
we obtain
\begin{align}
\e^{-2\i\vek{r}\cdot\vek{p}}\ps{V}^T(\vek{r},-\vek{p})& = 
\sum_{L, \Lambda'} (-\i)^{L+\Lambda'} (2L+1)\, j_L(2rp)\,\ps{V}^{T,\Lambda'}(r,p)
\sum_{\Lambda}
\ClebschGordan{L}{0}{\Lambda'}{0}{\Lambda}{0}^2
\frac{4\pi}{(2\Lambda+1)}\sum_{\mu} {Y^{\Lambda}_{\mu}}^*\!(\hat{r})\,{Y^{\Lambda}_{\mu}}(\hat{p})
\nonumber \\
& = \sum_{\Lambda}\sum_{\Lambda'}\ps{V}^{T,\Lambda'}(r,p)\ \left(
\sum_{L} (-\i)^{L+\Lambda'+\Lambda} (2L+1)\, j_L(2rp)\, 
\ClebschGordan{L}{0}{\Lambda'}{0}{\Lambda}{0}^2\, \right)
\i^\Lambda P_\Lambda(\hat{r}\ndot\hat{p})\\
& = \sum_{\Lambda}\sum_{\Lambda'}\ps{V}^{T,\Lambda'}(r,p)\, C_{\Lambda' \Lambda}(r,p)
\ \i^\Lambda P_\Lambda(\hat{r}\ndot\hat{p}) \, ,
\label{eq:exc2}
\end{align}
which defines the coefficients of Eq.~\eqref{eq:matrixeq2}
\begin{align}
C_{\Lambda' \Lambda}(r,p) = \sum_{L} (-\i)^{L+\Lambda'+\Lambda} (2L+1)\, j_L(2rp)\, 
\ClebschGordan{L}{0}{\Lambda'}{0}{\Lambda}{0}^2 \, .
\end{align}
\end{widetext}

\end{document}